\newcommand{\up}{\uparrow}
\newcommand{\down}{\downarrow}

\newcommand{\Id}[1] {\int \! \! {\rm d}^3 #1}

\newcommand{\vr} {{\bf r}}
\newcommand{\vR} {{\bf R}}

\newcommand{\ket}[1] { | #1 \rangle }

\documentclass[a4paper]{article}

\usepackage{epsfig}
\begin{document}

\title{
Coulomb couplings in positively charged fullerene
}


\author{Martin L{\"u}ders$^{1,2}$\thanks{E-mail: lueders@sissa.it},
        Andrea Bordoni$^{3}$,
        Nicola Manini$^{1,3,4}$\thanks{E-mail: nicola.manini@mi.infm.it},
\\
        Andrea Dal Corso$^{1,2}$,
        Michele Fabrizio$^{1,2,5}$\thanks{E-mail: fabrizio@sissa.it}
        and Erio Tosatti$^{1,2,5}$\thanks{E-mail: tosatti@sissa.it}
\\ \small \it 
$^1$ International School for Advanced Studies (SISSA),
Via Beirut 4, 34014 Trieste, Italy
\\ \small \it 
$^2$ INFM Democritos National Simulation Center, and INFM, Unit\`a
 Trieste, Italy
\\ \small \it 
$^3$ Dip.\ Fisica, Universit\`a di Milano, Via Celoria 16, 20133
Milano, Italy
\\ \small \it 
$^4$ INFM, Unit\`a di Milano, Milano, Italy
\\ \small \it 
$^5$ International Centre for Theoretical Physics (ICTP), 
P.O. Box 586, 34014 Trieste, Italy
}

\date{Sep 19, 2002}
\maketitle

\begin{abstract}
We compute, based on density-functional electronic-structure 
calculations, the Coulomb couplings in the $h_u$ highest occupied orbital of 
molecular C$_{60}$. We obtain a multiplet-averaged Hubbard 
$U\simeq 3$~eV, and four Hund-rule-like intra-molecular
multiplet-splitting terms, each of the order of few hundreds of
meVs. According to these couplings, all  C$_{60}^{n+}$ ions should
possess a high-spin ground state if kept in their rigid, undistorted form.
Even after molecular distortions are allowed, however,
the Coulomb terms still appear to be somewhat stronger than the previously 
calculated Jahn-Teller couplings, the latter favoring low-spin 
states. Thus for example in C$_{60}^{2+}$, unlike C$_{60}^{2-}$, 
the balance between Hund 
rule and Jahn Teller yields, even if marginally, a high-spin ground state. 
That seems surprising in view of reports of superconductivity in field-doped
C$_{60}^{n+}$ systems.

\end{abstract}

\section{Introduction}

Strong electron correlations in multi-band, orbitally
degenerate systems represent an important current theoretical challenge.
A lively experimental playground for that is provided by electron-doped
fullerene systems, which exhibit a variety of behavior, including
unconventional metals like cubic CsC$_{60}$ (Brouet {\it et al.}\ 1999),
superconductors
of the A$_3$C$_{60}$ family (A= K,Rb,Cs) (Ramirez 1994, Gunnarsson 1997), and
insulators, presumably of the Mott-Jahn-Teller
type (Fabrizio {\it et al.}\ 1997, Capone {\it et al.}\ 2000),
such as Na$_2$C$_{60}$ (Brouet {\it et al.}\ 2001),
A$_4$C$_{60}$ (Benning {\it et al.}\ 1993),
and the class of ammoniated compounds (NH$_3$)K$_{3-x}$Rb$_x$C$_{60}$.

The recently developed C$_{60}$ field effect transistor (FET) devices
(Sch\"on {\it et al.}\ 2000a,b) claimed metallic and superconducting states
for both electron and hole doping in the interface C$_{60}$ layer,
the hole-doped system showing generally higher $T_c$ than the
electron-doped system.  That could be
related to a larger electron-phonon (e-ph) coupling of the HOMO-derived
$h_u$ band than for the LUMO-derived $t_{1u}$ band (Manini {\it et al.}\ 2001).

However, in an orbitally degenerate system like the one at hand,
the electron-phonon coupling competes against intra-molecular 
exchange of Coulomb origin, responsible for Hund rules. In 
fact, Hund rules generally favor high spin for a degenerate 
molecular state, whereas coupling to intra-molecular vibrations 
leads to a Jahn-Teller (JT) splitting of the degeneracy which favors low spin.
Furthermore, since doped fullerenes are narrow-band molecular 
conductors, knowledge of the  local Coulomb repulsion, usually 
parametrized by the so-called Hubbard $U$, 
is important in order to establish whether C$_{60}^{n+}$ conductors
are weakly or strongly correlated electron systems. In the former
case, a conventional Eliashberg-type approach should be adequate
to explain superconductivity. In the latter, a new theoretical 
framework is most likely needed (Fabrizio {\it et al.}\ 1997, Capone {\it et
al.}\ 2000, Capone {\it et al.}\ 2002).   

All these considerations stress the importance of a realistic 
estimate  of the Coulomb interaction terms  
(Hubbard $U$ and the Hund multiplet terms) for C$_{60}^{n+}$.
In the past, electronic-structure-based calculations of these 
parameters have been made for the negative ions C$_{60}^{n-}$,
where the electrons are added to the $t_{1u}$ LUMO.
In that case the structure of the Coulomb Hamiltonian is formally the same
as that for an atomic $p$ level in spherical symmetry, and as such entirely
determined by two parameters only: the configuration-averaged $U$ and
Hund-rule intra-molecular exchange $J$ (Martin and Ritchie 1993, Han and
Gunnarsson 2000).
For positive ions C$_{60}^{n+}$, where $n$ holes are added to the $h_u$ HOMO
the only estimate available for the Coulomb parameters is a recent
empirical one (Nikolaev and Michel 2002).
The task of an electronic structure-based first principles calculation of
these parameters will be the main purpose of the present paper.

In a fivefold-degenerate $h_u$ orbital, as we shall detail below, icosahedral
symmetry determines these Coulomb couplings in terms of {\em five}
independent parameters: a configuration-averaged $U$, plus four
intra-molecular exchange terms.
All the low-energy electronic degrees of freedom of a solid-state system of
positively (or negatively) charged C$_{60}$ molecules can be well 
described by a model Hamiltonian including the five $h_u$ hole bands 
(or three $t_{1u}$ electron bands) only, all other orbitals 
acting, as usual, as a mere source of renormalization of the 
`bare' parameter values (Han and Gunnarsson 2000).
%

In order to calculate the five independent intra-molecular
electron-electron (e-e) Coulomb parameters, we use standard
density-functional electronic-structure calculations in the local
(spin) density approximation [L(S)DA],
imposing as a constraint different values 
of the electronic occupation number in the
individual Kohn-Sham (KS) orbitals.
We carry out several frozen-geometry single-molecule constrained-LDA
calculations of the total energy for a variety of charge 
and spin states of undistorted icosahedral C$_{60}^{n+}$. By 
comparing these energies
with the corresponding analytic expressions for the model Hamiltonian,
which is expressed in terms of the five unknown Coulomb parameters,
we finally determine all of them.

As the calculations are carried out for an isolated molecule, the computed
Coulomb parameters are effective values, which contain the screening due to
the polarizability of the filled molecular orbitals in the molecule, but
contain neither the screening due to the other molecules nor that 
of the conduction electrons in the solid.
As a reliability check, 
we also recompute with the same method the e-e $U$ and $J$ parameters
for the LUMO band. The results are found in good agreement with 
previous estimates 
(Martin and Ritchie 1993, Han and Gunnarsson 2000, Antropov {\it et al.}
1992), which further
confirms the viability of our method. 

With the Coulomb parameters in hand we can then compute the 
multiplet spectrum for any given molecular occupancy, $n$ =1,...5. 
This spectrum strictly applies only to ideal rigid C$_{60}$ ions,
and is not of direct experimental relevance, because it leaves
JT distortion effects out.
The latter can be crudely estimated using the hole-vibration couplings
previously calculated in C$_{60}$ (Manini {\it et al.}\ 2001) either at
second-order in perturbation theory, corresponding to a full neglect of
retardation effects, the so-called `anti-adiabatic' approximation, or in
the opposite `adiabatic' limit.
In the anti-adiabatic approximation, the effective e-e 
interaction is simply the superposition of the Coulomb repulsion 
and the phonon-mediated attraction.
The total net result as far as $U$ is concerned 
is still repulsive, the large Coulomb term only marginally
corrected by molecular distortions. All the
other intra-molecular exchange terms are instead heavily reduced. However,
while in the case of C$_{60}^{n-}$ that leads to an effective
sign reversal from Hund to `anti-Hund', (and from repulsive 
to attractive for an electron pair in the singlet channel) 
the balance is much less definite for C$_{60}^{n+}$,
where the overall sign remains positive for $n$=2 and is uncertain
for higher $n$ values.
The efficiency of the JT effect in reversing Hund-rules 
couplings for C$_{60}^{n+}$ is even weaker when the adiabatic approximation is 
considered instead of the anti-adiabatic one. In the adiabatic
approximation, where ionic motion is classical, the molecular ground 
state of C$_{60}^{n+}$ turns out to be always high spin for all $n$ 
values, in contrast to C$_{60}^{n-}$ where it is always low spin.     

This paper is organized as follows: Sec.~\ref{model:sec} introduces the
model, determining the minimal number of independent parameters consistent
with icosahedral symmetry.
The constrained-LDA calculation and its results are described in
Sec.~\ref{cLDA:sec}. 
The multiplet spectra resulting from the computed couplings are shown in
Sec.~\ref{multiplets:sec}.
The results are discussed in Sec.~\ref{discussion:sec}, and some lengthy
formulae are collected in an Appendix.

\section{The model Hamiltonian}

\label{model:sec}

Our final target is to address the low-energy properties of (a lattice of)
charged C$_{60}$ molecules.
To this end we construct a model Hamiltonian to describe the physics of
either the $h_u$ HOMO (holes) or $t_{1u}$ LUMO (electrons) bands.
The role of the other orbitals is to act as a renormalization of the
effective parameters for the band at the Fermi energy.
In this paper we concentrate on the determination of single-molecule
properties, and defer to a future work the calculation of the bands in the
solid.
The model Hamiltonian for a single molecule reads
\begin{equation}
\hat{H} = \hat{H}_0 + \hat{H}_{\rm vib}  + 
\hat{H}_{\rm e-vib} + \hat{H}_{\rm  e-e}
\label{modelhamiltonian}
\end{equation}
where 
\begin{eqnarray}
\hat{H}_0     &=& \epsilon \, \sum_{\sigma m}  
\hat{c}^\dagger_{\sigma m} \hat{c}_{\sigma m} \\
\label{vib-hamiltonian}
\hat{H}_{vib} &=& \sum_{i\Lambda \mu} \frac{\hbar \omega_{i\Lambda}}{2} 
( \hat{P}_{i\Lambda \mu}^2 + \hat{Q}_{i\Lambda \mu}^2 ) \\
\label{JT-hamiltonian}
\hat{H}_{\rm e-vib} &=& 
\sum_{r\,i\Lambda}
\frac{g^r_{i\Lambda} \hbar \omega_{i\Lambda}}{2} 
\sum_{\sigma m m' \mu} 
C^{r \Lambda \mu}_{m m'} \, \hat{Q}_{i\Lambda \mu} \,\hat{c}^\dagger_{\sigma m}
\hat{c}_{\sigma m'} \\
\hat{H}_{\rm  e-e} &=& 
\frac{1}{2} \sum_{\sigma, \sigma'} \sum_{{m m'}\atop{n n'}} 
w_{\sigma,\sigma'}(m,m';n,n') \,
\hat{c}^\dagger_{\sigma m} \hat{c}^\dagger_{\sigma' m'} \, 
\hat{c}_{\sigma' n'} \hat{c}_{\sigma n}.
\label{Coulomb-hamiltonian}
\end{eqnarray}
are respectively the single-particle Hamiltonian, the vibron contribution
(in the harmonic approximation), the electron-vibron coupling (in the
linear JT  approximation) (Manini {\it et al.}\ 2001,
Manini and De Los Rios 2000), and the mutual 
Coulomb repulsion between the electrons.
The $\hat{c}^\dagger_{\sigma, m}$ denote the creation operators of either
a hole in the HOMO or an electron in the LUMO, described by the
single-particle wave function $\varphi_{m\sigma}(\vr)$. 
$\sigma$ indicates the spin projection, $m$ and $n$ label the component
within the degenerate electronic HOMO/LUMO multiplet, and $i$ counts the
phonon modes of symmetry $\Lambda$ (2 $A_g$, 6 $G_g$ and 8 $H_h$ modes).
$C^{r \Lambda \mu}_{m n}$ are Clebsch-Gordan coefficients of the
icosahedron group, for coupling two $h_u$ (holes) or $t_{1u}$ (electrons)
states to phonons of symmetry $\Lambda$.
$r$ is a multiplicity label, relevant for $H_g$ modes only
(Manini {\it et al.}\ 2001, Butler 1981).
$\hat{Q}_{i\Lambda \mu}$ and $\hat{P}_{i\Lambda \mu}$ are the molecular
phonon coordinates and conjugate momenta.
Spin-orbit is exceedingly small (Tosatti {\it et al.}\ 1996),
and it is therefore neglected.

The Coulomb matrix elements are defined by:
\begin{equation}
\label{Coulomb-ints}
w_{\sigma,\sigma'}(m,m';n,n') = \Id{r} \! \Id{r'} \, 
\varphi^{*}_{m \sigma}(\vr) \, 
\varphi^{*}_{m'\sigma'}(\vr') \,
u_{\sigma,\sigma'}(\vr,\vr') \, 
\varphi_{n\sigma}(\vr) \,
\varphi_{n'\sigma'}(\vr') 
\end{equation}
where $u_{\sigma,\sigma'}(\vr,\vr')$ is an effective Coulomb repulsion,
screened by the other electrons of the molecule.

One way to estimate these matrix elements is to evaluate the
Coulomb integrals (\ref{Coulomb-ints}) directly for the simple kernel
$u_{\sigma,\sigma'}(\vr,\vr')= q_e^2 / (4\pi \epsilon_0 |\vr-\vr'|)$
and given molecular orbitals (Nikolaev and Michel 2002).
This approach neglects completely the screening due to the other electrons
on the same molecule.
Here we choose a rather different approach: namely, we parametrize the
interaction Hamiltonian (\ref{Coulomb-hamiltonian}) in the most general way
allowed by the molecular symmetry, and then determine the parameters by
fitting to {\em ab initio} electronic structure calculations.
As these calculations allow for the full polarization response of the total
charge density (except for core levels, whose polarizability is
negligible by comparison)
the screening effect of all molecular valence electrons is 
accounted for in the final parameters.

The symmetry of the Coulomb interaction plus the molecular symmetry of the
problem allow us to express all of the Coulomb integrals in
(\ref{Coulomb-ints}) as functions of a small set of physical parameters.
In the following, we obtain the minimal set of Coulomb parameters that
determine the interaction Hamiltonian (\ref{Coulomb-hamiltonian}), as
required by the symmetry of the molecule and the symmetry label of the
orbitals under consideration.

As the Hamiltonian is time-reversal invariant, 
the orbitals can be chosen real with no loss of generality.
Furthermore we take the orbitals, as well as the interaction, to be
spin-independent (thus neglecting spin-dependent screening effects
which might be possible in magnetic states), so that:
\begin{equation}
w_{\sigma,\sigma'}(m,m';n,n') = w(m,m';n,n') = \Id{r} \! \Id{r'} \, 
\varphi_{m}(\vr) \, 
\varphi_{m'}(\vr') \,
u(\vr,\vr') \, 
\varphi_{n}(\vr) \,
\varphi_{n'}(\vr') \ .
\end{equation}
With the above assumptions one finds immediately:
\begin{equation}
\label{symm-1}
w(m,m';n,n') = w(n,m';m,n') = w(m,n';n,m') = w(n,n';m,m') \ .
\end{equation}
The effective screened interaction shows the full molecular symmetry, i.e.
\begin{equation}
\label{symm-2}
u(\vR\, \vr,\vR\, \vr') = u(\vr,\vr')
\end{equation}
for all the symmetry operations $\vR$ of the icosahedral group $I_h$.
In order to make use of this symmetry, we decompose the product wave
functions into irreducible representations of the icosahedral group:
\begin{equation}
\varphi_{m}(\vr) \,
\varphi_{n}(\vr) =
\sum_{r, \Lambda, \mu} C^{r \Lambda \mu}_{m n} 
\Phi^{r \Lambda}_{\mu}(\vr) \ ,
\label{phicoupled:eq}
\end{equation}
using again the Clebsch-Gordan coefficients $C^{r \Lambda \mu}_{m n}$ to
couple two $h_u$ (holes) or $t_{1u}$ (electrons) tensors to an irreducible
tensor of symmetry $\Lambda$.
The label $\Lambda$ runs in principle on all the irreducible
representations ($A_g$, $T_{1g}$, $T_{2g}$, $G_g$, $H_g$, $A_u$, $T_{1u}$,
$T_{2u}$, $G_u$, $H_u$) of the icosahedral group $I_h$.
The multiplicity label $r$ distinguishes between multiple occurrences of
the same representation in the coupling (\ref{phicoupled:eq}): it is the
standard extra label for groups, such as $I_h$, which are not simply
reducible (Butler 1981).
Due to the symmetry relation (\ref{symm-1}), only the symmetric couplings
occur.
In particular, for holes in the HOMO, from the decomposition of
$h_u\otimes h_u$ the only nonzero contributions come from $\Lambda=A_g$,
$G_g$, $H_g^{(r=1)}$, and $H_g^{(r=2)}$.
For electrons in the $t_{1u}$ LUMO, we have $\Lambda=A_g$ and $H_g$ only.
In terms of this symmetry recoupling, we rewrite the Coulomb matrix
elements as:
\begin{equation}
w(m,m'; n, n')  = \! \! \! \sum_{ {r,\Lambda,\mu}\atop{r', \Lambda', \mu'} }
\! \! C^{r \Lambda \mu}_{m n}  \, 
C^{r' \Lambda' \mu'}_{m' n'} 
\! \Id{r} \Id{r'} \,
\Phi^{r \Lambda}_{\mu}(\vr) \,
u(\vr,\vr') \,
\Phi^{r' \Lambda'}_{\mu'}(\vr') \ .
\label{wdecomposed}
\end{equation}
This equation shows the decomposition of the $w(m,m'; n, n')$ interaction
matrix into the sum of products of geometric factors (Clebsch-Gordan
coefficients), times a relatively restricted number of coupled matrix
elements.
We can now exploit the symmetry of the Coulomb interaction (\ref{symm-2})
to further simplify the remaining integrals.
To this end, we apply a generic group operation $\vR$ to the integration 
variables.
The explicit transformation properties of the coupled wave functions
$\Phi^{r \Lambda}_{\mu}(\vr)$ allows to introduce the group
representation matrices $\Gamma^{\Lambda}_{\mu_1 \mu}(\vR)$, while the
effective interaction remains invariant.
Next, we apply the grand orthogonality theorem of representation theory,
to rewrite the interaction as follows:
\begin{eqnarray}
\lefteqn{
\Id{r} \Id{r'}
\Phi^{r \Lambda}_{\mu}(\vr) \,
u(\vr,\vr') \,
\Phi^{r' \Lambda'}_{\mu'}(\vr')}\nonumber \\
&=& 
\Id{r} \Id{r'}
\Phi^{r \Lambda}_{\mu}(\vR^{-1} \vr) \,
u(\vR^{-1} \vr,\vR^{-1} \vr') \,
\Phi^{r' \Lambda'}_{\mu'}(\vR^{-1}\vr')\nonumber \\
&=&
\sum_{\mu_1 \mu_1'} \Gamma^{\Lambda}_{\mu_1 \mu}(\vR)
\Gamma^{\Lambda'}_{\mu_1' \mu'}(\vR)
\Id{r} \Id{r'}
\Phi^{r \Lambda}_{\mu_1}(\vr) \,
u(\vr,\vr') \,
\Phi^{r' \Lambda'}_{\mu_1'}(\vr') \nonumber \\
&=&
\delta_{\Lambda, \Lambda'} \delta_{\mu,\mu'} 
\underbrace{\frac{1}{|\Lambda|} \sum_{\mu_1}  \Id{r} \Id{r'} 
\Phi^{r \Lambda}_{\mu_1}(\vr) \,
u(\vr,\vr') \,
\Phi^{r' \Lambda}_{\mu_1}(\vr') }_{F^{r,r',\Lambda}} .
\label{F-params}
\end{eqnarray}
This shows that the integrals in Eq.~(\ref{wdecomposed})
are diagonal in the representation label (but not in the multiplicity
label $r$).
This relation determines explicitly the most general expression for the
Coulomb matrix elements in a shell of $t_{1u}$ or $h_u$ icosahedral label,
in terms of a {\em minimal} set of independent parameters
$F^{r,r',\Lambda}$ [defined in Eq.~(\ref{F-params})]:
\begin{equation}
w(m,m';n,n') = \sum_{r,r',\Lambda} F^{r,r',\Lambda}
\left( \sum_\mu C^{r \Lambda \mu}_{m n} \, 
C^{r' \Lambda \mu}_{m' n'} \right) \ .
\label{Fdecomposition}
\end{equation}
In this paper we label states within the degenerate representation using
the $C_5$ quantum number $m$ from the $I_h\supset D_5\supset C_5$ group
chain (Butler 1981).
Note however that the purely geometric decomposition of the Coulomb
integrals (\ref{Fdecomposition}) holds for any choice of the group chain,
and correspondingly of the $I_h$ Clebsch-Gordan coefficients.

In the case of electron doping in the $t_{1u}$ orbital, no multiplicity $r$
labels appears, and thus, according to Eq.~(\ref{Fdecomposition}), the
Coulomb Hamiltonian is expressed as the sum of two terms, whose strength is
governed by the two parameters $F^{A_g}$ and $F^{H_g}$.
These parameters are related to the $k=0$ and $k=2$ Slater-Condon integrals
$F^{(k)}$ for $p$ electrons in spherical symmetry (Cowan 1981).
For hole doping in the $h_u$ HOMO, we need five parameters
\begin{equation}
F_1= F^{A_g},\
F_2= F^{G_g},\
F_3= F^{1,1,H_g},\
F_4= F^{2,2,H_g},\
F_5= F^{1,2,H_g}
\label{fnumbering}
\end{equation}
to determine completely the Coulomb matrix elements
\footnote{The Coulomb Hamiltonian for icosahedral $h$-states was expressed
in terms of five parameters also by Oliva 1997, Plakhutin and Carb\'o-Dorca
2000.}.
In terms of spherical symmetry for a $d$ atomic state, $F^{(0)}$ again
corresponds to the totally symmetric $F^{A_g}$ parameter, while the
$F^{(2)}$ and $F^{(4)}$ spherical parameters are replaced by the four
icosahedral $F_{2 \div 5}$.

Rather than the $F^\Lambda$ parameters, for $t_{1u}$ electrons it is more
common to use the parameters $U=F^{A_g}/3 - F^{H_g}/3$, and
$J= F^{H_g}/2$.
With this definition of $U$
\footnote{ It should be noted that $U$ differs from the usual definition
of the Hubbard $U$, involving the lowest multiplet in each
  $n$-configuration: $ U^{\rm min} = E^{\rm min}(n+1) + E^{\rm min}(n-1) -
  2 E^{\rm min}(n) $. This second definition is unconvenient, especially in
  the $h_u$ case, since it depends wildly on $n$.  },
the multiplet-averaged energy
has the simple dependence on the total number of electrons $n$:
\begin{equation}
E^{\rm ave}(n)=
{\rm Tr} ( H|_n )=
\epsilon n + U \frac {n (n-1)}2 \ ,
\end{equation}
where $H|_n$ is the Hamiltonian restricted to the $n$-electrons states.
The $J$ parameter controls the multiplet exchange splittings, so that the 
center of mass of the multiplets at fixed $n$ and total spin $S$ is 
given by \footnote{
In the $t_{1u}$ case the eigenenergies can be written in the closed form
$$
  E(n,S,L) = \epsilon \, n + U \frac{n (n-1)}{2} - 2 J \left[S(S+1) +
  \frac{1}{4}L(L+1) + \frac{1}{4}n(n-6)\right], $$ as function of $n$, $S$
  and
  the total `angular momentum' $L$ (recall that the $t_{1u}$ orbitals
  behave
  effectively as $p$-orbitals).}
\begin{equation}
E^{\rm ave}(n,S) =
{\rm Tr} ( H|_{n,S} )
= \epsilon \, n + U \frac{n (n-1)}{2} -
\frac{5}{4} J \left[S(S+1) - \frac 9{10} n +\frac{3}{20}n^2 \right] .
\label{JmultipletLUMO}
\end{equation}

For the $h_u$ holes, the center of mass of the multiplets with $n$ holes is
located at energy
\begin{equation}
E^{\rm ave}(n) = 
\epsilon \, n + \left( \frac{F_1}{5} - \frac{4 \, F_2}{45} 
- \frac{F_3}{9} - \frac{F_4}{9} \right) \frac{n (n-1)}{2}.
\end{equation}
This leads to the definition of an average Coulomb repulsion
\begin{equation}
U = \left( \frac{F_1}{5} - \frac{4 \, F_2}{45} 
- \frac{F_3}{9} - \frac{F_4}{9} \right) .
\label{Udefinition}
\end{equation}
We can define a spin-splitting parameter $J$ also for the holes, by
considering the center of mass of the multiplets at fixed spin $S$, $E^{\rm
ave}(n,S)$.
We find that the Coulomb Hamiltonian (\ref{Coulomb-hamiltonian}) 
is consistent with 
\begin{equation}
E^{\rm ave}(n,S) =
\epsilon \, n + U \frac{n (n-1)}{2} -
J \left[S(S+1) - \frac {5}{6} n +\frac{1}{12} n^2 \right] ,
\label{JmultipletHOMO}
\end{equation}
with
\begin{equation}
J=\frac 16 F_2 +\frac 5{24} \left(F_3 +F_4 \right) .
\label{Jdefinition}
\end{equation}
In what follows we take as a convenient set of independent 
Coulomb parameters: $U$, $F_2$, $F_3$, $F_4$, and $F_5$.

Finally, with the decomposition (\ref{Fdecomposition}) in hand, it is
convenient to re-organize the interaction Hamiltonian
(\ref{Coulomb-hamiltonian}) in terms of number-conserving symmetry-adapted
fermion operators:
\begin{equation}
\hat{H}_{\rm e-e} = \frac{1}{2} \sum_{r r' \Lambda} F^{r,r',\Lambda} 
\left( \sum_\mu \hat{w}^{r \Lambda \mu} \, \hat{w}^{r' \Lambda \mu} \right)
- A \hat{n} 
\label{He-ecombined}
\end{equation}
where we defined the operators:
\begin{equation}
\hat{w}^{r \Lambda \mu} := \sum_{\sigma} \sum_{m n} 
C^{r \Lambda \mu}_{m n} \, \hat{c}^\dagger_{\sigma m} \hat{c}_{\sigma n}
\end{equation}
and the constant $A$, which is
$\left( \frac{1}{2} U + \frac{8}{3} J \right)$ for holes and
$\left( \frac{1}{2} U + 2 J \right)$ for electrons.


\section{Determination of the Coulomb parameters}

\label{cLDA:sec}

After the explicit derivation of the form of a general icosahedral e-e
interaction $\hat{H}_{\rm e-e}$, we come now to the numerical calculation
of the parameters fixing the interaction for C$_{60}$ ions.
We compute these parameters by comparing the (analytical) expressions for
the energies in the model Hamiltonian (\ref{modelhamiltonian}) with
numerical results, obtained by first-principles density-functional theory
(DFT) LDA calculations of the electronic structure of the C$_{60}$
molecule.

As the Coulomb Hamiltonian governs the spectrum of multiplet excited states
of the ionized configurations, in principle it would be straightforward
to obtain the Coulomb parameters by fitting the excitation energies of
(\ref{modelhamiltonian}) to multiplet energies obtained with some {\it
ab-initio} method,
or to experimental multiplet spectra, if they were available
and if the e-ph contribution could be separated out.
However, to our knowledge no such experimental data are till now available.
We thus choose to extract the Coulomb parameters from DFT calculations.
Yet this is not straightforward, since in standard DFT the excitation
energies of the KS system do not have a rigorous physical
meaning; the KS states are only auxiliary quantities (Perdew 1985).
Although excitation energies are accessible in DFT within the framework of
time-dependent DFT (Petersilka {\it et al.}\ 1996),
here we follow an alternative approach
which is similar in spirit to the constrained-LDA
(Dederichs {\it et al.}\ 1984, Gunnarsson {\it et al.}\ 1989)
method to extract effective local Coulomb parameters.
In a nutshell, constrained-LDA yields the ground-state energy of the
system subject to some external constraint, such as a fixed magnetization
or a fixed orbital occupancy.
In practice, it is convenient to impose constraints such as to select states
which are single Slater determinants, since they are described fairly
accurately by standard LDA methods.
By comparing their total energies with the expectation values of the model
Hamiltonian with respect to the same states, it is possible to determine
the interaction parameters, in the spirit of the $\Delta$SCF
(self-consistent field) scheme (Jones and Gunnarsson 1989).

%
%

In order to describe the method, it is convenient to consider 
a simple example. Let us focus on the states 
$|n_\up,n_\down;0,0;0,0\rangle$ in the LUMO subspace where 
$n_\up$ spin up and $n_\down$ spin down electrons fill   
orbital $\varphi_1$, the other two orbitals being empty, and 
define $E^{\rm tot}(n_\up,n_\down)$ the corresponding total energies.
One could determine the `Hubbard $U_1$' 
relating to this orbital as
\begin{equation}
U_1 = E^{\rm tot}(1,1)
+E^{\rm tot}(0,0) - E^{\rm tot}(1,0) - E^{\rm tot}(0,1) 
= E^{\rm tot}(1,1) + E^{\rm tot}(0,0) - 
2E^{\rm tot}(1,0) \ .
\label{Uoneorbital}
\end{equation}
A slight complication to this simple approach is brought in our problem by
the orbital degeneracy.
Suppose we indeed compute by LDA the energy of $|1,1;0,0;0,0\rangle$.
This orbital, where we place the two electrons, is initially degenerate to
the other two LUMO components.
However, in LDA, a Kohn-Sham (KS) filled orbital shifts immediately up in
energy with respect to empty ones.
Consequently, if we insist to fill one of the three, originally degenerate,
KS orbitals, at the next iteration the electrons go naturally to occupy
either of the two initially empty orbitals, with a large jump in charge
density from one iteration to the next.
This effect is due to the imperfect cancellation of the self interaction
within LDA
\footnote{ Self-interaction is not the only possible origin of the
convergence problem.
  The subtle problem of pure-state $v$-representability leads to basically
  the
  same symptoms. See (Schipper {\it et al.}\ 1998).},
which is also related to the well-known LDA 
gap problem.
%
%
%
%

A possible remedy is to artificially introduce small gaps (of the order of
10~meV) in the otherwise degenerate HOMO and LUMO, by adding a tiny
distortion of the icosahedral C$_{60}$ molecule along one of the JT active
modes.
Since the distortion is very small (each atom moving by less than 0.5~pm
from its equilibrium position), the model remains essentially
representative of fully symmetric fullerene.
However, in order to effectively control the charge (either electrons in
the LUMO or holes in the HOMO) in the different orbitals, the splittings
must overcome the self-repulsion.
This fact suggests occupying an orbital by a small fraction of an electron
so that the self repulsion is sufficiently small to leave
this orbital in the same energy position dictated by the distortion field.

To recover the actual value of the energy at integer charge, as required
for the determination of $U$ according to Eq.~(\ref{Uoneorbital}), we make
use of a known artifact of the LSDA:
the ground-state energy of a system as a functional of the fractional
occupation of a KS orbital interpolates smoothly the energies
of integer multiples of the elementary charge.
To the extent that the modifications of orbital $\phi_1$ 
due to changes of its
filling $n_1$ may be neglected, the total energy is a parabolic function of
charge. In particular, in our specific example, 
\begin{eqnarray}
E^{\rm tot}\left(\frac{n_1}{2},\frac{n_1}{2}\right) 
&=& E^{\rm tot}_0 + b \, n_1 + \frac{c}{2} \, n_1^2 \ ,\nonumber\\
E^{\rm tot}\left(n_1,0\right) 
&=& E^{\rm tot}_0 + b' \, n_1 + \frac{c'}{2} \, n_1^2 \ ,
\label{parabola}
\end{eqnarray}
where, according to Janak's theorem (Janak 1978), the linear coefficients 
$b=b'$ equal the KS single-particle energy $\epsilon_1$ at $n_1\to 0$.
The quadratic coefficients $c$ and $c'$ are not generally the same, since
they are extracted from unpolarized and spin-polarized configurations,
respectively.
We have verified, in those configurations where the self interaction causes
no convergence problem, that the extrapolation from calculations at $n_1\leq
0.2$ is in very good quantitative agreement with direct total-energy
calculations at integer $n_1$'s.
Through Eqs. (\ref{parabola}) and (\ref{Uoneorbital}) we have:
\begin{equation}
U_1= \left(E^{\rm tot}_0 + b\cdot 2 + \frac{c}{2}\cdot 2^2 \right) 
+ E^{\rm tot}_0 -2\left( E^{\rm tot}_0+ b' +
\frac{c'}{2} 1^2\right)=  (2 c -c') \ .
\label{Uexample}
\end{equation}
We have therefore expressed the Hubbard $U_1$ as a linear combination of
the quadratic coefficients $c$ and $c'$ of the extrapolation parabolas.
Note that the $c$'s for {\em both} configurations involved are needed for
the determination of $U_1$: it would be incorrect to identify $U_1$ to, for
example, the curvature $c$ of the total LDA energy as a function of charge.

\begin{figure}[ht]
\centerline{
\epsfig{file=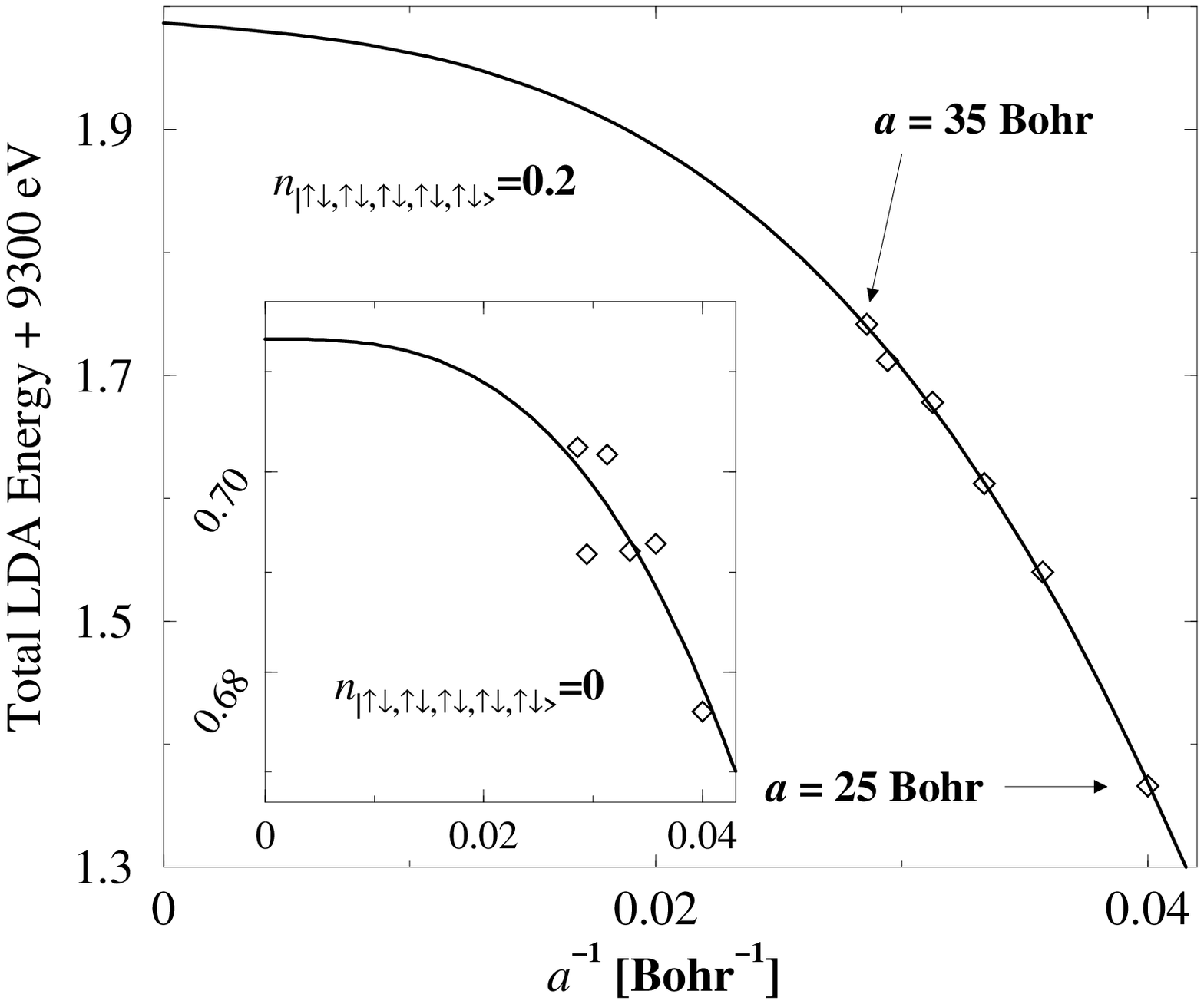,width=10.0cm}
}
\caption{
The total LDA energy (diamonds) of the C$_{60}$ molecule as a function of
the inverse lattice parameter $a^{-1}$ for (uniformly spread) hole charge
$n_{\ket{\up \down, \up \down, \up \down, \up \down, \up \down}}=0.2$ (main
graph) and $n_{\ket{\up \down, \up \down, \up \down, \up \down, \up
\down}}=0$ (inset).
The solid lines are finite-size polynomial scalings, including a
Madelung $a^{-1}$ term, fixed in accordance to (Makov and Payne 1995), plus 
constant and  $a^{-3}$ fitted terms.
\label{Energia.lda:fig}}
\end{figure}

The complete determination of the two and respectively of the five 
Coulomb parameters for $t_{1u}$ electrons and for $h_u$ holes follows 
the same track as the simple determination of
$U_1$ outlined above.
First, we select two sets of electronic configurations, one for electron
and the other for hole doping, containing three (see
Table~\ref{c60-coefs:tab}) and eight (see Table \ref{c60+coefs:tab})
elements, respectively.
For each configuration $|i\rangle$, we compute the total energy as a 
function of the
(fractional) charge, for five values $0\leq n_i \leq 0.2$.
The calculations for a given set are carried out with a fixed JT 
distortion within DFT-LSDA.
As in previous calculations (Manini {\it et al.}\ 2001) we use ultrasoft 
pseudopotentials (Vanderbilt 1990) for C (Favot and Dal Corso 1999).
The plane-waves basis set is cut off at 27~Ry (charge
density cutoff = 162~Ry).
The C$_{60}$ molecule is repeated periodically in a large simple-cubic
supercell lattice of side $a$.
To insure total charge neutrality, thus correcting for the $G=0$ divergence
of the total energy, a compensating uniform background charge is added.
The total energy is corrected for the leading power-law Coulomb
interactions among supercells, by removing the Madelung $a^{-1}$ term and
the $a^{-3}$ correction, with the method devised by Makov and
Payne (1995).
We extract the finite-$a$ corrections by running several
calculations with $a$ ranging between 1.32 and 1.85~nm, as illustrated in
Fig.~\ref{Energia.lda:fig} in a typical example.
[It might have been marginally cheaper to use the modified Coulomb
potential method (Jarvis {\it et al.}\ 1997) instead of the size scaling.
That method however required a larger lattice parameter $a$,
and thus more memory space].
Parabolas of the form (\ref{parabola}) are fitted to the calculated
$a\to\infty$ energies.
The resulting quadratic coefficients $c_i$ for electrons and holes are
reported in Tables~\ref{c60-coefs:tab} and \ref{c60+coefs:tab}
respectively.

In the light of Janak's theorem (Janak 1978), stating that the
single-particle KS levels $\epsilon_i(n_i)=\partial E^{\rm
tot}(n_i)/\partial n_i$, the $c_i$ coefficients, besides representing
second derivatives of the total energy w.r.t.\ charge, can alternatively be
seen as first derivatives of the single-particle levels w.r.t.\ charge, as
follows:
\begin{equation}
\epsilon_i(n_i) = \epsilon_i(0) + c_i \, n_i +O(n_i^2) \ .
\label{linear}
\end{equation}
  From the $a$-scaling of the total energy (Makov and Payne 1995),
we derive the
$a$-scaling of the single-particle KS levels $\epsilon_i$, which allows us
to compute these quantities for the isolated molecular ion ($a\to \infty$).
Equation~(\ref{linear}) (neglecting $O(n_i^2)$ corrections) provides a
second method to derive the $c_i$ coefficients, which are the basic
ingredients in the calculation of the Coulomb parameters.
As apparent in Tables~\ref{c60-coefs:tab} and \ref{c60+coefs:tab}, the
coefficients $c_i$ obtained from the total energy and from the
single-particle levels are essentially in accord.
However, the values from the single-particle levels are numerically more
stable since, contrary to the total-energy method, they do not involve
small differences of large numbers.
In the following we shall use the $c_i$'s from single-particle energies for
the determination of the Coulomb parameters.

The calculation of the e-e parameters is then realized by equating the
various DFT extrapolated energies to the expectation values of
(\ref{modelhamiltonian}) with respect to the same electronic 
configurations.
Given the arbitrariness in the reference energy  
$\epsilon$ for the model Hamiltonian (\ref{modelhamiltonian}),   
we define for convenience  
\begin{equation}
\epsilon = U/2 + b_i - \mu \ .
\label{herecomesmu}
\end{equation}
where $\mu$ is a free parameter allowing for the possibility of a chemical
potential shift with respect to the DFT calculation \footnote{Notice that
$\mu$ is not exactly a chemical potential shift, since the linear
  term within our constrained DFT-LSDA has a component proportional to the
  Jahn-Teller splitting introduced to stabilize each set of electronic
  configurations.}.

In Table~\ref{comparison-:table} we collect the analytic expression of the
energies of the three states considered for $t_{1u}$ electrons.
Equating the terms on the third and fourth column of
Table~\ref{comparison-:table}, we have 3 equations to determine the 3
unknown quantities $U$, $J$ and $\mu$: we obtain the physical parameters
simply by inversion of the linear dependency, and by replacing the values
of $c_i$ in Table~\ref{c60-coefs:tab}.
We have tabulated the combination $2 n_i^{-2}\left(\langle i|\hat{H}_{\rm
e-e}|i\rangle + U n_i /2\right)$ instead of simply $\langle i|\hat{H}_{\rm
e-e}|i\rangle$, as each equation involves quantities of the same order of
magnitude. 
By eliminating $\mu$ [in analogy to the one-state example of
Eq.~(\ref{Uexample})], we find for the Coulomb parameters of the negative
C$_{60}$ ions: $U=\frac 65 c_{\ket{\up \down, \up \down, \up \down}} -
\frac 15 c_{\ket{0 , \up, 0}} = 3069$~meV and $J= \frac 65
c_{\ket{\up \down, \up \down, \up \down}} - \frac 32 c_{\ket{\up, \up, \up}}
 + \frac 3{10} c_{\ket{0 , \up, 0}} = 32$~meV.
These values, summarized in Table~\ref{parameters-:table}, are in the same
range as previous estimates (Martin and Ritchie 1993, Han and Gunnarsson
2000, Antropov {\it et al.}\ 1992).

To produce a reliable estimate of the six unknown quantities (the five e-e
parameters plus $\mu$), we consider eight different $h_u$ hole states, for
whose energies we collect the analytic expressions in
Table~\ref{comparison+:table}.
Therefore we have eight equations in six unknowns: we obtain the best
estimate of the physical parameters by adjusting them to minimize the sum
of the squared difference between the energies in the third and the fourth
column of Table~\ref{comparison+:table}.
For this overdetermined system of equations, the combination $2
n_i^{-2}\left(\langle i|\hat{H}_{\rm e-e}|i\rangle + U n_i /2\right)$ shows
its advantage, that all LSDA calculations weight the same in the fit.
This fitting procedure yields the values of the Coulomb parameters for
C$_{60}^{n+}$ collected in Table~\ref{parameters+:table}.
The standard deviation of the fit (1.4~meV) 
gives an estimate of the numerical accuracy of the $c_i$ parameters.
By standard error propagation, we obtain the estimate of the errorbar on
the individual e-e parameters reported in Table~\ref{parameters+:table}.

We can now comment on our obtained results.
We observe first of all that the only large parameter is $U$.
It takes essentially the same value in the LUMO and the HOMO: this value of
about 3~eV governs the multiplet-averaged hole-hole repulsion, and is
also compatible with experimental estimates (Antropov {\it et al.}\ 1992)
for isolated molecular ions.
In the solid, the screening of the local Coulomb parameters due to the
polarizability of all the surrounding C$_{60}$ molecules could be
approximately accounted for in a Clausius-Mossotti scheme
(Antropov {\it et al.}\ 1992),
and, for C$_{60}^{n-}$, it may reduce $U$ by roughly a
factor 0.5 (Lof {\it et al.}\ 1992).
The polarizability screening in the solid is expected to affect $F_2\div
F_5$ much less than $U$.
Note however that the actual Hubbard $U$, based on differences of
ground-state energies, acquires an $n$-dependent
contribution of $F_2\div F_5$.
The appropriate $U^{\rm min}$ are collected in Table~\ref{Umintable}.
The extra intra-molecular contribution is especially large at half filling.

The relative smallness of the intra-molecular parameters compared to $U$ is
traced to the very close values of the quadratic coefficients $c_i$
(listed in Table~\ref{c60+coefs:tab}) for all different configurations
$|i\rangle$.
In turn, this indicates that, contrary to the strongly
localized orbitals of atomic physics, in C$_{60}$ it does not matter much
the relative spin and orbital placement of two electrons in the HOMO or
LUMO: they would always feel more or less the same repulsion of roughly
3~eV.
The Hund rules are therefore rather weak in C$_{60}$ ions, because 
the degenerate orbitals are spread over a carbon shell of 7~\AA
diameter, rather than concentrated around a single nucleus.
The largest parameter is $F_3$ corresponding to the $r=1$ $H_g$ symmetry.
The $F_5$ parameter is the smallest, effectively compatible within 
error bars with a zero value.

%
%
The computed intra-molecular exchange $J$ is almost twice as large
\footnote{ The definition of $J$ for the $t_{1u}$ orbital contains a
factor 5/4
  due to historical reasons, as apparent from the comparison of
  Eq.~(\ref{JmultipletLUMO}) and (\ref{JmultipletHOMO}). If this factor is
  accounted for, the actual ratio between the first-Hund-rule terms in
  the
  HOMO and in the LUMO is about 1.5}
 in the HOMO than in the LUMO
(Tables~\ref{parameters-:table} and \ref{parameters+:table}).
Consequently, the splittings (in the order of hundreds of meV) of the
$n$-hole Coulomb multiplets are larger for holes than for electrons.
We come next to the detailed study of this multiplet spectrum.


\section{Multiplet energies}

\label{multiplets:sec}

The computed values of the coupling parameters can be used to calculate the
multiplet spectrum of $\hat{H}_{\rm e-e}$.
%
%
We concentrate here on the states of $n$ holes in the fivefold
degenerate $h_u$ HOMO.
In order to diagonalize the Hamiltonian matrix, we wish to 
take full advantage of symmetry.  For each charge $n$, 
$(I_h\supset D_5)$ orbital label $(\Lambda,
\mu$), multiplicity $r$, total spin $S$, and spin projection $M$, we first
construct a set of symmetry-adapted states by iteratively coupling the
one-hole $h_u$ orbitals to all the $(n-1)$-holes states as follows:
\begin{eqnarray}
\ket{n, (\tau', \Lambda', S'), (\Lambda, r), \mu; S, M} &=& 
\sum_{\mu' \mu''} \sum_{M' M''}
C^{r \Lambda \mu}_{\Lambda', \mu'; h_u, \mu''} \,
C^{S M}_{S', M'; 1/2, M''} \, \times \nonumber \\
&&
\ket{n-1, \tau', \Lambda', \mu'; S', M'} \, \times
\ket{1, h_u, \mu''; \frac 12, M''}
\end{eqnarray}
where $C^{r \Lambda \mu}_{\Lambda', \mu'; h_u, \mu''}$ and $C^{S M}_{S',
M'; 1/2, M''}$ are the icosahedral and spherical Clebsch-Gordan
coefficients taking care of the orbital and spin recoupling respectively,
and $\tau'$ is the parentage of the $(n-1)$-particle state.
The resulting states $\ket{n, (\tau', \Lambda', S'), (\Lambda, r), \mu; S,
M}$ (for all possible $\tau', \Lambda', S'$ and $r$) are then
orthonormalized to form the set of $n$-hole basis states
$\ket{n,\tau,\Lambda,\mu; S, M}$, with $\tau$ counting their parentage.

In this symmetry-adapted basis the Hamiltonian is diagonal
with respect to the labels $n, \Lambda, \mu, S$ and $M$
and its eigenvalues are independent of $\mu$ and $M$.
This block-diagonal form allows in many cases to compute the analytical
expressions for the multiplet energies $E_{\rm mult}$, i.e.\ the
eigenvalues of $\hat{H}_{\rm e-e} -U\hat{n}(\hat{n}-1)/2$, that are 
collected in Table~\ref{analyticMultiplet}.
For $3\leq n\leq 7$, the calculation of the multiplet energies involves the
diagonalization of block matrices of size 3 up to 7, where analytical 
methods are unpractical.
For these cases we report the Hamiltonian submatrices for those states in
Appendix~\ref{matrices}.
The analytical 3-holes spectrum shows the degeneracy of a $T_{1u}$ and a
$T_{2u}$ doublet state, which has been observed and explained in
previous work (Oliva 1997, Lo and Judd 1999, Plakhutin and Carb{\'o}-Dorca
2000).

The spectrum obtained by substituting the computed parameters of
Table~\ref{parameters+:table} into the expressions of
Table~\ref{analyticMultiplet} is collected in Table~\ref{numericMultiplet}.
For all values of $n$, we of course verify Hund-rule behavior, 
i.e.\ the high-spin
state has the lowest energy. However the first Hund rule leads to comparable
splittings to second Hund rule, so that states of different spin are
energetically inter-mixed, for $n=4$ and 5.
The multiplet structure is similar to what was reported in
Ref. (Nikolaev and Michel 2002),
with a few differences in the detailed ordering of
closely spaced levels.
The total spread in the DFT results of Table~\ref{numericMultiplet},
however, are a factor three smaller than those of that rigid-orbital
unscreened calculation (Nikolaev and Michel 2002).

The computed spectrum of Table~\ref{numericMultiplet} represents that of
ideal rigid icosahedral C$_{60}$.
The coupling of electronic state to the intramolecular vibrations 
generally leads to JT distortions, involving energy scales
that compete with the e-e repulsion and generally favor low-spin states,
a sort of anti-Hund rule.
The interplay of the Coulomb and e-ph terms originates a complex pattern of
vibronic multiplet states that was studied in detail in the simpler case of
the negative C$_{60}$ ions (O'Brien 1996).
The HOMO system at hand is much more intricate, due to the interplay of
several parameters.
Here we shall address this problem at a more approximate level of accuracy.

First we observe that, in the limit where the typical phonon energies
$\hbar\omega_i$ are much larger than the e-ph energy gains $\sim g_i^2
\hbar\omega_i$, the e-ph Hamiltonian treated at second order in $g_i$
takes the form of the first term in the right hand side of
Eq.~(\ref{He-ecombined}) ({\em anti-adiabatic} or weak-coupling limit).
The strengths of the {\em effective} e-e interaction parameters are given
in terms of the dimensionless couplings $g^r_{i\Lambda}$ and frequency
$\omega_{i\Lambda}$ by
$\tilde F^{r,r',\Lambda} = 
- \sum_i g^r_{i\Lambda}g^{r'}_{i\Lambda} \hbar \omega_{i\Lambda}/4
$ 
\footnote{ According to (Manini {\it et al.}\ 2001), the coefficients for the
coupling of $H_g$
  with $H_g$ to $A_g$ and to $G_g$ are normalized so that $\sum_{n m}
  \left(C^{A_g\,\mu}_{n m}\right)^2= 5$ and $\sum_{n m}
  \left(C^{G_g\,\mu}_{n
  m}\right)^2= 5/4$. Here we prefer to apply the standard normalization to
  unity, thus we include the 5 and 5/4 factors into the $g^2$ coupling
  parameters. Within this convention, a factor 6 must be included in $g^2$
  for
  the coupling of $t_{1u}$ electrons to $H_g$ modes.}.
Table~\ref{parameters+:table} lists the parameters $\tilde F_i$ generated
by the e-ph, in the notation of Eqs.~(\ref{fnumbering}) (\ref{Udefinition})
and (\ref{Jdefinition}), compared to the $F_i$ Coulomb parameters.
At this level of approximation, the e-e and e-ph terms are expressed as
sums of formally identical terms, differing only for the value of the
parameter multiplying each term.
Thus, it is natural to combine the two contributions into a {\em total}
effective two-body Hamiltonian $\hat{H}_{\rm e-e}^{\rm tot}$ which is 
formally identical to the definition of
Eqs.~(\ref{Coulomb-hamiltonian}) and (\ref{Fdecomposition}) 
but based on total effective e-e parameters given by the
algebraic sum of the Coulomb and e-ph parameters.
These total effective e-e parameters are listed in the
last column of Table~\ref{parameters+:table}.

%

The contributions to $U$ of the $G_g$ and $H_g$ phonons
[Eq.~(\ref{Udefinition})] is larger than the $A_g$ term, thus giving a
small repulsive phononic multiplet-averaged $\tilde U$.
However, the huge $F_1$ Coulomb `monopole' parameter is barely affected
by the $A_g$ phonons-originated term.
Accordingly, the total multiplet-averaged interaction $U^{\rm tot}$ is left
basically unaffected by the phonons contribution, for both the HOMO and the
LUMO.

\begin{figure}[ht]
\centerline{
\epsfig{file=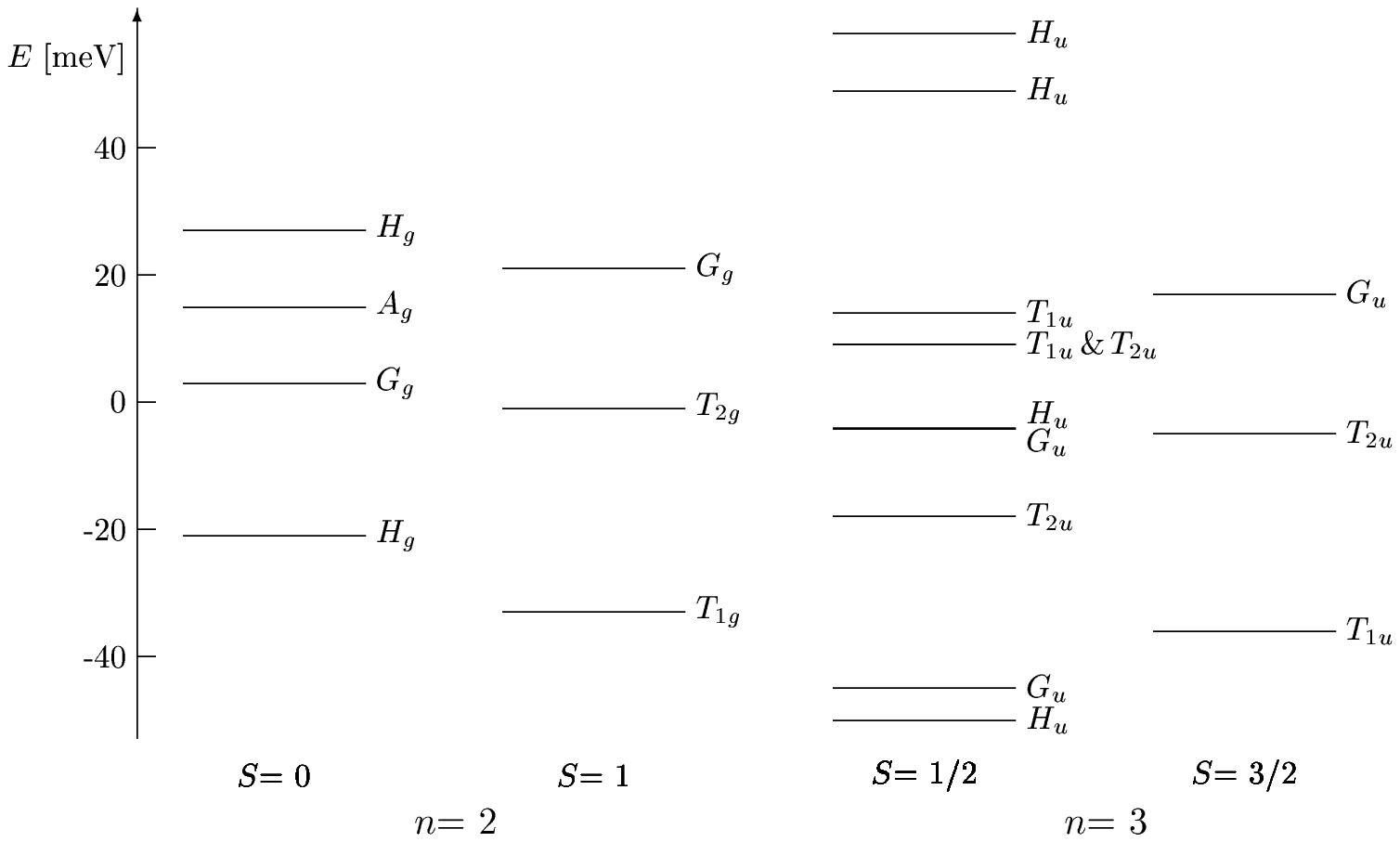,width=12.0cm}
}
\caption{
The multiplet spectra for $n=2$ and 3 holes, including both the e-e and
the e-ph couplings in the anti-adiabatic approximation, as given by the
total effective parameters listed in the last column of
Table~\ref{parameters+:table}.
The low-spin ground state for $n\geq 3$ is probably an artifact of the
anti-adiabatic overestimation of e-ph energetics.
\label{multipletSpectrumTotal:fig}}
\end{figure}

On the contrary, the intra-molecular Hund terms $F_i$ are of values
comparable to the phononic counterpart, thus leading to a strong
cancellation.
In particular, for the holes the largest repulsive term $F_3$ is reversed
by the even larger coupling $\tilde F_3$ to the $H_g^{(r=1)}$ phonon modes.
The largest total effective term is $F_2$: it remains positive, due to the
modest $G_g$-phonons--mediated attraction.
The corresponding multiplets spectrum, is reported in
Table~\ref{effectiveMultiplet}, and drawn for $n=2$ and $n=3$ in
Fig.~\ref{multipletSpectrumTotal:fig}.
For $n=2$ holes the ground state remains a triplet, with a very small gap
to the lowest singlet, while low-spin ground states prevail for $n\geq 3$.
For electrons, the total effective $J=-25$~meV indicates that low-spin
anti-Hund states are to be expected for C$_{60}^{n-}$, as is indeed
observed experimentally in many electron-doped C$_{60}$ compounds
(Kiefl {\it et al.}\ 1992, Zimmer {\it et al.}\ 1995, Lukyanchuk  {\it et
al.}\ 1995, Prassides  {\it et al.}\ 1999).

\begin{figure}[ht]
\centerline{
\epsfig{file=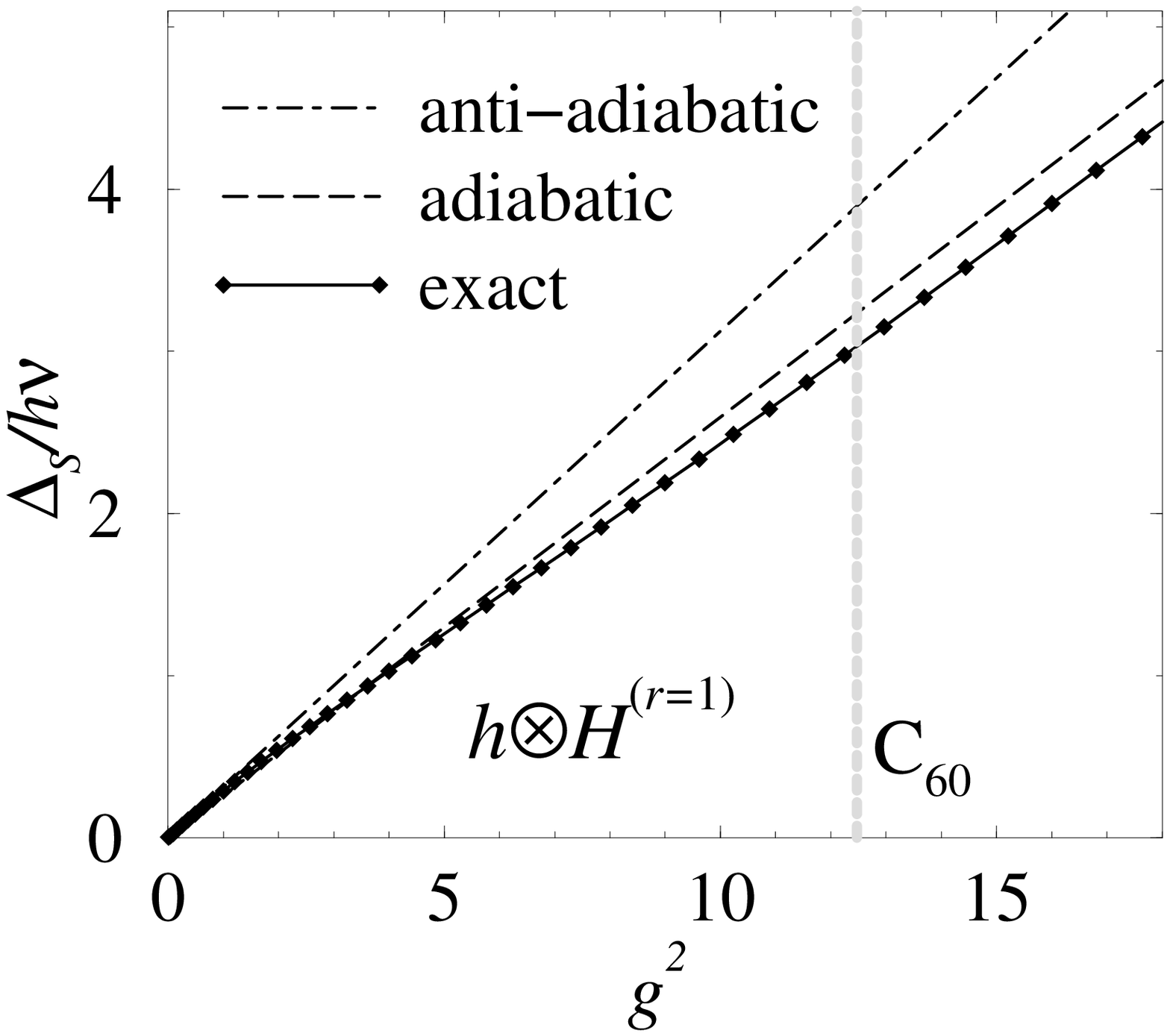,width=10.0cm}
}
\caption{
The $n=2$ holes JT spin gap to the lowest triplet state, as a function of
the dimensionless linear coupling strength $g^2=-4 \tilde
F_3/(\hbar\omega)$ of a single $H_g$ mode (O'Brien 1972, Manini and Tosatti
1998) (no Coulomb
interaction) with pure $r=1$ coupling, similar to the 271~cm$^{-1}$ mode of
C$_{60}$, the one with the largest coupling.
The same spin gap is plotted for comparison in the anti-adiabatic
(dot-dashed -- slope 5/16) and adiabatic (dashed -- slope 83/320) limits.
%
The vertical bar at $g^2=12.5$ locates the total effective $r=1$ coupling
of C$_{60}$ positive ions, according to the DFT estimate of
Manini {\it et al.}\ (2001).
\label{hbyHJT:fig}}
\end{figure}

The anti-adiabatic approximation, although tending to overestimate
the e-ph energies is advantageous in allowing to map the
e-ph Hamiltonian onto an effective attractive e-e term: this mapping does
not apply any longer when the coupling energies are not taken as much
smaller than the harmonic phonon energies.
However, the relatively large values of the realistic 
e-ph coupling in both positive and negative C$_{60}$ ions make the 
weak-coupling approximation not truly justified. Indeed the e-ph energetics 
based on this approximation are grossly overestimated if applied to
intermediate/strong couplings.
In practice, the JT energy gains and gaps in units of $g_i^2\hbar\omega_i$
become significantly smaller as the coupling changes from weak ($g_i\ll 1$)
to strong ($g_i\gg 1$).
In particular, for $t_{1u}$ electrons interacting only with $H_g$ modes (no
e-e terms), the energy lowering divided by $g_i^2\hbar\omega_i$ drops by
60~\% from weak to strong coupling (Manini  {\it et al.}\ 1994).
For $h_u$ holes, for $n=2$ and a single $H_g^{(r=1)}$ mode
(O'Brien 1972, Manini and Tosatti 1998), we see in Fig.~\ref{hbyHJT:fig} that 
the spin gap in units of $g^2\hbar\omega$ reduces by 17~\% only, going from
the weak to the strong-coupling limit.
However, when all C$_{60}$ modes are included, this reduction is as large
as $\sim$50~\% due to contributions of the $H_g^{(r=2)}$ and $G_g$ modes.

As the actual (anti-Hund) e-ph coupling should have weaker effects than
those estimated in the anti-adiabatic approximation, the question of what
is the symmetry of the ground state of the C$_{60}^{n\pm}$ ions remains
open.
The case of C$_{60}^{2+}$ marks an exception, since Coulomb couplings
prevail already at the anti-adiabatic level
(Fig.~\ref{multipletSpectrumTotal:fig}). Thus the prediction of an $S=1$
magnetic ground state for $n$=2 holes in C$_{60}$ seems fairly robust,
at least within LDA accuracy.
In order to settle this problem for the other cases, we study the e-ph
coupling in the opposite, adiabatic limit, which becomes exact in the limit
of strong e-ph coupling, and which proved quantitatively more realistic for
C$_{60}^{n\pm}$ ions (Fig.~\ref{hbyHJT:fig}).
At the adiabatic level, the phonons are treated classically,
with the electrons/holes contributing through (\ref{JT-hamiltonian}) to the
total adiabatic potential acting on the phonon coordinates
$\hat{Q}_{i\Lambda \mu}$ which are treated as classical variables.
The additional ingredient we include here, and which was not included in
the previous adiabatic calculation (Manini  {\it et al.}\ 2001)
is the e-e coupling.
In the JT-distorted configuration, the icosahedral symmetry is broken,
therefore all symmetry states are mixed.
Only $n$, $S$ and $S_z$ are conserved.
For example, the coupling to the distortion of $n=3$ holes, $S=3/2$, $M=3/2$ 
mixes the 10 states of $T_{1u}$, $T_{2u}$ and $T_{2u}$ symmetry listed in
Table~\ref{analyticMultiplet}.
Assuming that the Coulomb parameters $F_i$ are unchanged upon distortion,
for each $n$, $S$ and $M$ we determine the optimal distortion, by full
minimization of the lowest adiabatic potential sheet in the space of all
the phonons coordinates.
In Table~\ref{adiabatic_energies}, we report the resulting lowest-state
energy in each spin sector, based on the e-e and e-ph couplings of
C$_{60}^{n\pm}$ ions.
As already stated, we find a difference between electrons and holes:
the ground state is low spin for C$_{60}^{2-}$ electrons, while it is
always high spin for C$_{60}^{n+}$.
The case of C$_{60}^{3-}$ has almost degenerate $S=1/2$ and $S=3/2$ states,
the former probably prevailing once non-adiabatic corrections are 
accounted for (Capone  {\it et al.}\ 2001).
For the positive ions instead, the adiabatic result overturns the
anti-adiabatic prediction of low-spin ground state for $n\geq 3$ holes.
In these cases, the adiabatic spin gaps are fairly large, and are likely 
to survive when zero-point quantum corrections to the adiabatic 
approximations are added.
For C$_{60}^{2+}$ the gap to the lowest singlet state is rather small, but,
as noted above, here the ground state is a spin triplet even in the
anti-adiabatic approximation: it is likely to remain high spin also within
an exact treatment of the phonons.

The outcome of the adiabatic calculation is that 
positive C$_{60}$ ions favor high-spin ground states, while in negative ions
e-ph coupling prevails and low-spin ground states are likely.
However, for C$_{60}^{n\pm}$ ions the balance of e-e and e-ph is rather
delicate, therefore the problem of the spin symmetry of the ground state
ions is far from trivial, and remains basically open.
Indeed, in some chemical environments high-spin states are observed to
prevail in negative fullerene ions (Schilder {\it et al.}\ 1994, Arovas and
Auerbach 1995), and this
indicates that the lowest multiplets of different spin type are almost
degenerate.
To get a more conclusive answer on this point for ions of both signs, it
would be crucial to carry out a full diagonalization of
(\ref{modelhamiltonian}) including all the phonon modes and Coulomb terms
on the same ground, on the line of O'Brien (1996): we plan to carry out
such calculation in a future work.


\section{Discussion and Conclusions}
\label{discussion:sec}

The Coulomb couplings of holes in C$_{60}$ obtained in this paper 
are based on rigid icosahedral geometry calculations.
However, clearly in each different-charge state, the C$_{60}^{n\pm}$
molecular ion relaxes to different equilibrium positions, according to the
interplay of e-ph coupling ($\hat{H}_{\rm vib} +\hat{H}_{\rm e-vib}$) with
intra-molecular Coulomb exchange ($\hat{H}_{\rm e-e}$).
In principle one could compute the Coulomb parameters, allowing
simultaneously for geometry relaxation.
The disadvantage of such a calculation is the difficulty of
disentangling the e-ph and e-e contributions.
A second difficulty of principle is that the
ion, in an electronically degenerate state, distorts to several equivalent
static JT minima of less than icosahedral symmetry (Manini and De Los Rios
2000).
These local minima are connected by tunneling matrix elements which mix
them to suitable dynamical combinations of the different distortions, thus
restoring the original icosahedral symmetry: such non-adiabatic situation
would be outside the range of applicability of current standard
first-principles computational methods, usually based on the
Born-Oppenheimer separation of the ionic and electronic motions.
Moreover, the lack of exact cancellation of self-interaction
in the LDA makes even a practical attempt at a static, adiabatic
calculation for JT-distorted ions impossible at this stage. 
These are the reasons that suggested restricting this first Coulomb
calculation to the rigidly undistorted geometry.

A further limitation of the present calculation is the assumption that
the Coulomb parameters are independent of the charge of the state.
In principle, due to both orbital and geometrical relaxation, the effective
Coulomb interaction (\ref{Coulomb-ints}) will depend on the instantaneous
charge state of the fullerene ion.
However, this effect, a very important one in single-atom calculations, is
expected to be fairly small in such a large molecule as C$_{60}$.
Thus our parameters represent an average over $n$.

For C$_{60}^{n-}$, the value of $J$ was estimated in the 100~meV region by
Hartree-Fock calculations (Chang {\it et al.}\ 1991), and by direct
integration of the unscreened Coulomb kernel (Nikolaev and Michel 2002).
Both these methods overestimate the Coulomb repulsion because of
underestimation or complete neglect of screening.
The LSDA, where we get $J=32$~meV is on the other hand known
to overestimate screening, and thus to underestimate the Coulomb parameters.
Some value in between, such as $J\sim 50$~meV, as suggested by 
Martin and Ritchie (1993) is probably a more realistic
estimate of $J$ in C$_{60}^{n-}$.
Coming to the C$_{60}^{n+}$ case, we can regard the
e-e couplings of Table~\ref{parameters+:table} as lower bounds, the actual
repulsion being possibly a factor 1.5 or 2 larger.
Indeed, the calculation of Nikolaev and Michel (2002) finds splittings about 3
times larger than those of Table~\ref{numericMultiplet}, and those can 
reasonably be regarded as upper bounds.

On the other hand, the competing e-ph interaction is also very likely
underestimated by LDA, as was demonstrated in the case of C$_{60}^-$ by the
comparison of the calculated $g^2$ couplings to those extracted 
from fitting the
photoemission spectrum (Gunnarsson {\it et al.}\ 1995), which suggested values 
roughly twice as large.
In that case, the effective e-ph $\tilde J$, determined from 
the photoemission data is
$\tilde J=-127$~meV compared to the LDA value of $\tilde J=-57$~meV.
In conclusion, both the Coulomb repulsion and the phonon-mediated 
attraction calculated within LDA are likely to need a
rescaling by a similar factor of order two. Thus the balance 
between the two opposing interactions remains delicate in 
both C$_{60}^{n-}$ [as demonstrated by the presence of both 
high-spin and low-spin local ground states in different chemical environments
(Brouet {\it et al.}\ 2001, Kiefl {\it et al.}\ 1992, Zimmer {\it et al.}
1995, Lukyanchuk {\it et al.}\ 1995, Prassides  {\it et al.}\ 1999, Schilder
{\it et al.}\ 1994, Arovas and Auerbach 1995)]
and even more so in C$_{60}^{n+}$, where however the high-spin states 
should be more favored.
Moreover, especially in the hole doped case,  
we find multiplet splittings which are 
comparable to the theoretical bandwidth of
solid-state fullerene (about 0.5~eV), indicating that Hund-rule
intramolecular interactions are an important ingredient in C$_{60}$ ions.
%
%

Since any treatment of superconductivity caused by the JT 
coupling must include the competing Hund-rule terms, our results 
surprisingly indicate that positively-doped C$_{60}$ could display
a {\em weaker} tendency toward superconductivity than negatively-doped
C$_{60}$.  
Magnetic states could occur for any integer hole filling,
more commonly than for integer electron fillings. Even if magnetism
were to be removed owing to band effects, one should still
expect C$_{60}^{n-}$ to make better superconductors.
This conclusion is unexpected in the light of recent data claiming
a larger superconducting $T_c$ in positively charged than in negatively 
charged C$_{60}$ FETs (Sch\"on {\it et al.}\ 2000a,b). The reasons for this
disagreement are presently unclear, and will require further 
theoretical and experimental work.

\section*{Acknowledgments}

We are indebted to O.\ Gunnarsson, G.\ Onida and G.\ Santoro
for useful discussions.
This work was supported by the European Union, contract
ERBFMRXCT970155 (TMR FULPROP), covering in particular the 
postdoctoral work of M. Lueders, and by MURST COFIN01. 
The calculations were carried out using the PWSCF package (Baroni {\it et
al.}\ 2002) and were made possible by a ``Grant promozionale di
supercalcolo'' by INFM and CINECA.

\appendix 
\section{Appendix}
\label{matrices}

We provide here the recipe to construct the matrices $H_{\rm
mult}[n,\Lambda,S]$, whose eigenvalues $E_{\rm mult}$ give the $(F_2 \dots
F_5)$-dependent contribution to the multiplet energies of $n$ holes of
global symmetry $\Lambda$ and total spin $S$ in
Table~\ref{analyticMultiplet}.
Each matrix $H_{\rm mult}[n,\Lambda,S]$ is a linear combination of four
numerical matrices $M_i[n,\Lambda,S]$ (given below), with as coefficients
the Coulomb parameters $F_2\dots F_5$:
\begin{equation}
H_{\rm mult}[n,\Lambda,S] = 
\sum_{i=2}^5 \, F_i \, \, M_i[n,\Lambda,S] \ .
\label{matrixdecomposition}
\end{equation}
With specific choice of the parameters, it is possible to study the effect
of one particular operator: for example, by taking only $F_3\neq 0$, thus
diagonalizing the $M_3[n,\Lambda,S]$ matrices one may study analytically
the multiplet spectrum associated to that operator.
To get the spectra of Tables \ref{numericMultiplet} and
\ref{effectiveMultiplet}, we plugged the parameters of
Table~\ref{parameters+:table} into Eq. (\ref{matrixdecomposition}), and
proceeded to diagonalize $H_{\rm mult}[n,\Lambda,S]$ numerically.

The matrices $M_i[n,\Lambda,S]$ are as follows:
\begin{eqnarray*}
M_{2}[3,H_u,1/2] &=&  \left( \matrix{  
\frac{2}{3} &  
-\frac{2}{{\sqrt{21}}} &  
{\sqrt{\frac{10}{21}}} &  
0 \cr 
-\frac{2}{{\sqrt{21}}} &  
- \frac{2}{21}   &  
-\frac{5\,{\sqrt{\frac{5}{2}}}}{21} &  
-{\sqrt{\frac{5}{42}}} \cr 
{\sqrt{\frac{10}{21}}} &  
-\frac{5\,{\sqrt{\frac{5}{2}}}}{21} &  
\frac{11}{42} &  
- \frac{1}{{\sqrt{21}}}   \cr 
0 &  
-{\sqrt{\frac{5}{42}}} &  
- \frac{1}{{\sqrt{21}}}   &  
\frac{5}{6}
 }  \right) \\ 
M_{3}[3,H_u,1/2] &=&  \left( \matrix{  
\frac{5}{6} &  
0 &  
0 &  
0 \cr 
0 &  
\frac{5}{42} &  
\frac{3\,{\sqrt{\frac{5}{2}}}}{14} &  
\frac{{\sqrt{\frac{15}{14}}}}{2} \cr 
0 &  
\frac{3\,{\sqrt{\frac{5}{2}}}}{14} &  
- \frac{17}{84}   &  
\frac{{\sqrt{\frac{3}{7}}}}{2} \cr 
0 &  
\frac{{\sqrt{\frac{15}{14}}}}{2} &  
\frac{{\sqrt{\frac{3}{7}}}}{2} &  
\frac{1}{12}
 }  \right) \\ 
M_{4}[3,H_u,1/2] &=&  \left( \matrix{  
\frac{5}{6} &  
\frac{2}{{\sqrt{21}}} &  
-{\sqrt{\frac{10}{21}}} &  
0 \cr 
\frac{2}{{\sqrt{21}}} &  
\frac{13}{42} &  
\frac{{\sqrt{\frac{5}{2}}}}{42} &  
-\frac{{\sqrt{\frac{5}{42}}}}{2} \cr 
-{\sqrt{\frac{10}{21}}} &  
\frac{{\sqrt{\frac{5}{2}}}}{42} &  
\frac{23}{84} &  
-\frac{1}{2\,{\sqrt{21}}} \cr 
0 &  
-\frac{{\sqrt{\frac{5}{42}}}}{2} &  
-\frac{1}{2\,{\sqrt{21}}} &  
- \frac{7}{12}  
 }  \right) \\ 
M_{5}[3,H_u,1/2] &=&  \left( \matrix{  
0 &  
{\sqrt{\frac{15}{7}}} &  
{\sqrt{\frac{6}{7}}} &  
0 \cr 
{\sqrt{\frac{15}{7}}} &  
\frac{2\,{\sqrt{5}}}{7} &  
-\frac{3}{7\,{\sqrt{2}}} &  
- \frac{1}{{\sqrt{42}}}   \cr 
{\sqrt{\frac{6}{7}}} &  
-\frac{3}{7\,{\sqrt{2}}} &  
-\frac{2\,{\sqrt{5}}}{7} &  
\frac{{\sqrt{\frac{5}{21}}}}{2} \cr 
0 &  
- \frac{1}{{\sqrt{42}}}   &  
\frac{{\sqrt{\frac{5}{21}}}}{2} &  
0
 }  \right) \\ 
\end{eqnarray*}
\begin{eqnarray*}
M_{2}[4,A_g,0] &=&  \left( \matrix{  
\frac{4}{3} &  
-\frac{4}{{\sqrt{21}}} &  
2\,{\sqrt{\frac{10}{21}}} \cr 
-\frac{4}{{\sqrt{21}}} &  
- \frac{4}{21}   &  
-\frac{5\,{\sqrt{10}}}{21} \cr 
2\,{\sqrt{\frac{10}{21}}} &  
-\frac{5\,{\sqrt{10}}}{21} &  
\frac{11}{21}
 }  \right) \\ 
M_{3}[4,A_g,0] &=&  \left( \matrix{  
\frac{5}{3} &  
0 &  
0 \cr 
0 &  
\frac{5}{21} &  
\frac{3\,{\sqrt{\frac{5}{2}}}}{7} \cr 
0 &  
\frac{3\,{\sqrt{\frac{5}{2}}}}{7} &  
- \frac{17}{42}  
 }  \right) \\ 
M_{4}[4,A_g,0] &=&  \left( \matrix{  
\frac{5}{3} &  
\frac{4}{{\sqrt{21}}} &  
-2\,{\sqrt{\frac{10}{21}}} \cr 
\frac{4}{{\sqrt{21}}} &  
\frac{13}{21} &  
\frac{{\sqrt{\frac{5}{2}}}}{21} \cr 
-2\,{\sqrt{\frac{10}{21}}} &  
\frac{{\sqrt{\frac{5}{2}}}}{21} &  
\frac{23}{42}
 }  \right) \\ 
M_{5}[4,A_g,0] &=&  \left( \matrix{  
0 &  
2\,{\sqrt{\frac{15}{7}}} &  
2\,{\sqrt{\frac{6}{7}}} \cr 
2\,{\sqrt{\frac{15}{7}}} &  
\frac{4\,{\sqrt{5}}}{7} &  
-\frac{3\,{\sqrt{2}}}{7} \cr 
2\,{\sqrt{\frac{6}{7}}} &  
-\frac{3\,{\sqrt{2}}}{7} &  
-\frac{4\,{\sqrt{5}}}{7}
 }  \right) \\ 
\end{eqnarray*}
\begin{eqnarray*}
M_{2}[4,G_g,0] &=&  \left( \matrix{  
- \frac{3}{8}   &  
-\frac{1}{16\,{\sqrt{3}}} &  
- \frac{11}{48}   &  
\frac{1}{{\sqrt{6}}} \cr 
-\frac{1}{16\,{\sqrt{3}}} &  
\frac{25}{96} &  
\frac{1}{32\,{\sqrt{3}}} &  
-\frac{1}{2\,{\sqrt{2}}} \cr 
- \frac{11}{48}   &  
\frac{1}{32\,{\sqrt{3}}} &  
\frac{13}{160} &  
\frac{1}{10\,{\sqrt{6}}} \cr 
\frac{1}{{\sqrt{6}}} &  
-\frac{1}{2\,{\sqrt{2}}} &  
\frac{1}{10\,{\sqrt{6}}} &  
\frac{11}{5}
 }  \right) \\ 
M_{3}[4,G_g,0] &=&  \left( \matrix{  
\frac{23}{48} &  
-\frac{11\,{\sqrt{3}}}{32} &  
- \frac{9}{32}   &  
-\frac{{\sqrt{\frac{3}{2}}}}{2} \cr 
-\frac{11\,{\sqrt{3}}}{32} &  
\frac{65}{192} &  
-\frac{13\,{\sqrt{3}}}{64} &  
-\frac{1}{4\,{\sqrt{2}}} \cr 
- \frac{9}{32}   &  
-\frac{13\,{\sqrt{3}}}{64} &  
\frac{83}{192} &  
-\frac{{\sqrt{\frac{3}{2}}}}{4} \cr 
-\frac{{\sqrt{\frac{3}{2}}}}{2} &  
-\frac{1}{4\,{\sqrt{2}}} &  
-\frac{{\sqrt{\frac{3}{2}}}}{4} &  
\frac{1}{6}
 }  \right) \\ 
M_{4}[4,G_g,0] &=&  \left( \matrix{  
\frac{9}{16} &  
\frac{35}{32\,{\sqrt{3}}} &  
\frac{49}{96} &  
\frac{1}{2\,{\sqrt{6}}} \cr 
\frac{35}{32\,{\sqrt{3}}} &  
\frac{133}{192} &  
\frac{61}{64\,{\sqrt{3}}} &  
-\frac{1}{4\,{\sqrt{2}}} \cr 
\frac{49}{96} &  
\frac{61}{64\,{\sqrt{3}}} &  
\frac{73}{320} &  
\frac{1}{20\,{\sqrt{6}}} \cr 
\frac{1}{2\,{\sqrt{6}}} &  
-\frac{1}{4\,{\sqrt{2}}} &  
\frac{1}{20\,{\sqrt{6}}} &  
- \frac{9}{10}  
 }  \right) \\ 
M_{5}[4,G_g,0] &=&  \left( \matrix{  
\frac{{\sqrt{5}}}{8} &  
-\frac{5\,{\sqrt{\frac{5}{3}}}}{16} &  
-\frac{17}{16\,{\sqrt{5}}} &  
\frac{1}{{\sqrt{30}}} \cr 
-\frac{5\,{\sqrt{\frac{5}{3}}}}{16} &  
-\frac{13\,{\sqrt{5}}}{32} &  
\frac{149}{32\,{\sqrt{15}}} &  
\frac{1}{2\,{\sqrt{10}}} \cr 
-\frac{17}{16\,{\sqrt{5}}} &  
\frac{149}{32\,{\sqrt{15}}} &  
\frac{9\,{\sqrt{5}}}{32} &  
-\frac{{\sqrt{\frac{5}{6}}}}{2} \cr 
\frac{1}{{\sqrt{30}}} &  
\frac{1}{2\,{\sqrt{10}}} &  
-\frac{{\sqrt{\frac{5}{6}}}}{2} &  
0
 }  \right) \\ 
\end{eqnarray*}
\begin{eqnarray*}
M_{2}[4,H_g,0] &=&  \left( \matrix{  
\frac{32}{63} &  
-\frac{17\,{\sqrt{\frac{5}{11}}}}{21} &  
\frac{2\,{\sqrt{\frac{5}{1771}}}}{9} &  
-\frac{7}{3\,{\sqrt{69}}} &  
-\frac{{\sqrt{2}}}{3} \cr 
-\frac{17\,{\sqrt{\frac{5}{11}}}}{21} &  
\frac{29}{462} &  
-\frac{67}{33\,{\sqrt{161}}} &  
-2\,{\sqrt{\frac{15}{253}}} &  
{\sqrt{\frac{5}{22}}} \cr 
\frac{2\,{\sqrt{\frac{5}{1771}}}}{9} &  
-\frac{67}{33\,{\sqrt{161}}} &  
- \frac{2269}{4554}   &  
\frac{146\,{\sqrt{\frac{5}{231}}}}{69} &  
-\frac{17\,{\sqrt{\frac{5}{3542}}}}{3} \cr 
-\frac{7}{3\,{\sqrt{69}}} &  
-2\,{\sqrt{\frac{15}{253}}} &  
\frac{146\,{\sqrt{\frac{5}{231}}}}{69} &  
\frac{218}{483} &  
-\frac{16\,{\sqrt{\frac{2}{69}}}}{7} \cr 
-\frac{{\sqrt{2}}}{3} &  
{\sqrt{\frac{5}{22}}} &  
-\frac{17\,{\sqrt{\frac{5}{3542}}}}{3} &  
-\frac{16\,{\sqrt{\frac{2}{69}}}}{7} &  
\frac{1}{7}
 }  \right) \\ 
M_{3}[4,H_g,0] &=&  \left( \matrix{  
\frac{5}{12} &  
-\frac{{\sqrt{\frac{5}{11}}}}{2} &  
\frac{13\,{\sqrt{\frac{5}{1771}}}}{2} &  
-\frac{4\,{\sqrt{\frac{3}{23}}}}{7} &  
\frac{5}{14\,{\sqrt{2}}} \cr 
-\frac{{\sqrt{\frac{5}{11}}}}{2} &  
\frac{43}{132} &  
-\frac{29}{22\,{\sqrt{161}}} &  
\frac{{\sqrt{\frac{165}{23}}}}{14} &  
-\frac{13\,{\sqrt{\frac{5}{22}}}}{14} \cr 
\frac{13\,{\sqrt{\frac{5}{1771}}}}{2} &  
-\frac{29}{22\,{\sqrt{161}}} &  
\frac{749}{3036} &  
-\frac{{\sqrt{\frac{165}{7}}}}{46} &  
-\frac{{\sqrt{\frac{5}{3542}}}}{2} \cr 
-\frac{4\,{\sqrt{\frac{3}{23}}}}{7} &  
\frac{{\sqrt{\frac{165}{23}}}}{14} &  
-\frac{{\sqrt{\frac{165}{7}}}}{46} &  
\frac{2413}{1932} &  
\frac{13\,{\sqrt{\frac{3}{46}}}}{14} \cr 
\frac{5}{14\,{\sqrt{2}}} &  
-\frac{13\,{\sqrt{\frac{5}{22}}}}{14} &  
-\frac{{\sqrt{\frac{5}{3542}}}}{2} &  
\frac{13\,{\sqrt{\frac{3}{46}}}}{14} &  
\frac{25}{42}
 }  \right) \\ 
M_{4}[4,H_g,0] &=&  \left( \matrix{  
\frac{61}{252} &  
\frac{5\,{\sqrt{55}}}{42} &  
\frac{{\sqrt{\frac{385}{23}}}}{18} &  
\frac{40}{21\,{\sqrt{69}}} &  
-\frac{5}{42\,{\sqrt{2}}} \cr 
\frac{5\,{\sqrt{55}}}{42} &  
\frac{425}{924} &  
\frac{47\,{\sqrt{\frac{7}{23}}}}{66} &  
\frac{43\,{\sqrt{\frac{15}{253}}}}{14} &  
-\frac{5\,{\sqrt{\frac{5}{22}}}}{14} \cr 
\frac{{\sqrt{\frac{385}{23}}}}{18} &  
\frac{47\,{\sqrt{\frac{7}{23}}}}{66} &  
\frac{15419}{9108} &  
-\frac{25\,{\sqrt{\frac{35}{33}}}}{138} &  
-\frac{5\,{\sqrt{\frac{35}{506}}}}{6} \cr 
\frac{40}{21\,{\sqrt{69}}} &  
\frac{43\,{\sqrt{\frac{15}{253}}}}{14} &  
-\frac{25\,{\sqrt{\frac{35}{33}}}}{138} &  
\frac{433}{1932} &  
-\frac{5}{14\,{\sqrt{138}}} \cr 
-\frac{5}{42\,{\sqrt{2}}} &  
-\frac{5\,{\sqrt{\frac{5}{22}}}}{14} &  
-\frac{5\,{\sqrt{\frac{35}{506}}}}{6} &  
-\frac{5}{14\,{\sqrt{138}}} &  
\frac{3}{14}
 }  \right) \\ 
M_{5}[4,H_g,0] &=&  \left( \matrix{  
-\frac{11\,{\sqrt{5}}}{42} &  
-\frac{65}{21\,{\sqrt{11}}} &  
-\frac{71}{3\,{\sqrt{1771}}} &  
-\frac{4\,{\sqrt{\frac{5}{69}}}}{7} &  
-\frac{{\sqrt{\frac{5}{2}}}}{21} \cr 
-\frac{65}{21\,{\sqrt{11}}} &  
-\frac{95\,{\sqrt{5}}}{154} &  
\frac{25\,{\sqrt{\frac{5}{161}}}}{33} &  
-\frac{95}{7\,{\sqrt{759}}} &  
\frac{37}{7\,{\sqrt{22}}} \cr 
-\frac{71}{3\,{\sqrt{1771}}} &  
\frac{25\,{\sqrt{\frac{5}{161}}}}{33} &  
\frac{1301\,{\sqrt{5}}}{1518} &  
-\frac{17\,{\sqrt{\frac{7}{33}}}}{23} &  
-\frac{29\,{\sqrt{\frac{7}{506}}}}{3} \cr 
-\frac{4\,{\sqrt{\frac{5}{69}}}}{7} &  
-\frac{95}{7\,{\sqrt{759}}} &  
-\frac{17\,{\sqrt{\frac{7}{33}}}}{23} &  
\frac{{\sqrt{5}}}{46} &  
11\,{\sqrt{\frac{5}{138}}} \cr 
-\frac{{\sqrt{\frac{5}{2}}}}{21} &  
\frac{37}{7\,{\sqrt{22}}} &  
-\frac{29\,{\sqrt{\frac{7}{506}}}}{3} &  
11\,{\sqrt{\frac{5}{138}}} &  
0
 }  \right) \\ 
\end{eqnarray*}
\begin{eqnarray*}
M_{2}[4,T_{1g},1] &=&  \left( \matrix{  
\frac{8}{133} &  
-\frac{13\,{\sqrt{\frac{5}{57}}}}{7} &  
-\frac{22\,{\sqrt{\frac{2}{7}}}}{19} \cr 
-\frac{13\,{\sqrt{\frac{5}{57}}}}{7} &  
\frac{1}{14} &  
5\,{\sqrt{\frac{5}{798}}} \cr 
-\frac{22\,{\sqrt{\frac{2}{7}}}}{19} &  
5\,{\sqrt{\frac{5}{798}}} &  
\frac{2}{57}
 }  \right) \\ 
M_{3}[4,T_{1g},1] &=&  \left( \matrix{  
- \frac{85}{1596}   &  
\frac{5\,{\sqrt{\frac{15}{19}}}}{14} &  
-\frac{25}{38\,{\sqrt{14}}} \cr 
\frac{5\,{\sqrt{\frac{15}{19}}}}{14} &  
\frac{17}{84} &  
\frac{{\sqrt{\frac{15}{266}}}}{2} \cr 
-\frac{25}{38\,{\sqrt{14}}} &  
\frac{{\sqrt{\frac{15}{266}}}}{2} &  
\frac{20}{57}
 }  \right) \\ 
M_{4}[4,T_{1g},1] &=&  \left( \matrix{  
- \frac{83}{532}   &  
\frac{11\,{\sqrt{\frac{5}{57}}}}{14} &  
\frac{43}{38\,{\sqrt{14}}} \cr 
\frac{11\,{\sqrt{\frac{5}{57}}}}{14} &  
- \frac{17}{28}   &  
-\frac{13\,{\sqrt{\frac{5}{798}}}}{2} \cr 
\frac{43}{38\,{\sqrt{14}}} &  
-\frac{13\,{\sqrt{\frac{5}{798}}}}{2} &  
\frac{34}{57}
 }  \right) \\ 
M_{5}[4,T_{1g},1] &=&  \left( \matrix{  
-\frac{41\,{\sqrt{5}}}{266} &  
-\frac{29}{7\,{\sqrt{57}}} &  
\frac{51\,{\sqrt{\frac{5}{14}}}}{19} \cr 
-\frac{29}{7\,{\sqrt{57}}} &  
-\frac{3\,{\sqrt{5}}}{14} &  
\frac{17}{{\sqrt{798}}} \cr 
\frac{51\,{\sqrt{\frac{5}{14}}}}{19} &  
\frac{17}{{\sqrt{798}}} &  
\frac{7\,{\sqrt{5}}}{19}
 }  \right) \\ 
\end{eqnarray*}
\begin{eqnarray*}
M_{2}[4,T_{2g},1] &=&  \left( \matrix{  
- \frac{7}{58}   &  
-\frac{16\,{\sqrt{\frac{10}{39}}}}{29} &  
\frac{23}{{\sqrt{1131}}} \cr 
-\frac{16\,{\sqrt{\frac{10}{39}}}}{29} &  
- \frac{371}{1131}   &  
-\frac{5\,{\sqrt{\frac{10}{29}}}}{13} \cr 
\frac{23}{{\sqrt{1131}}} &  
-\frac{5\,{\sqrt{\frac{10}{29}}}}{13} &  
\frac{8}{13}
 }  \right) \\ 
M_{3}[4,T_{2g},1] &=&  \left( \matrix{  
\frac{5}{12} &  
0 &  
0 \cr 
0 &  
- \frac{11}{78}   &  
\frac{{\sqrt{\frac{145}{2}}}}{26} \cr 
0 &  
\frac{{\sqrt{\frac{145}{2}}}}{26} &  
\frac{35}{156}
 }  \right) \\ 
M_{4}[4,T_{2g},1] &=&  \left( \matrix{  
- \frac{49}{116}   &  
\frac{31\,{\sqrt{\frac{10}{39}}}}{29} &  
-\frac{8}{{\sqrt{1131}}} \cr 
\frac{31\,{\sqrt{\frac{10}{39}}}}{29} &  
\frac{1057}{2262} &  
\frac{41\,{\sqrt{\frac{5}{58}}}}{26} \cr 
-\frac{8}{{\sqrt{1131}}} &  
\frac{41\,{\sqrt{\frac{5}{58}}}}{26} &  
- \frac{11}{52}  
 }  \right) \\ 
M_{5}[4,T_{2g},1] &=&  \left( \matrix{  
-\frac{17\,{\sqrt{5}}}{58} &  
-\frac{157\,{\sqrt{\frac{2}{39}}}}{29} &  
-8\,{\sqrt{\frac{5}{1131}}} \cr 
-\frac{157\,{\sqrt{\frac{2}{39}}}}{29} &  
-\frac{136\,{\sqrt{5}}}{377} &  
\frac{59}{13\,{\sqrt{58}}} \cr 
-8\,{\sqrt{\frac{5}{1131}}} &  
\frac{59}{13\,{\sqrt{58}}} &  
\frac{17\,{\sqrt{5}}}{26}
 }  \right) \\ 
\end{eqnarray*}
\begin{eqnarray*}
M_{2}[4,G_g,1] &=&  \left( \matrix{  
\frac{77}{232} &  
\frac{25\,{\sqrt{\frac{15}{11}}}}{464} &  
\frac{497}{48\,{\sqrt{319}}} \cr 
\frac{25\,{\sqrt{\frac{15}{11}}}}{464} &  
- \frac{16979}{30624}   &  
\frac{425\,{\sqrt{\frac{5}{87}}}}{352} \cr 
\frac{497}{48\,{\sqrt{319}}} &  
\frac{425\,{\sqrt{\frac{5}{87}}}}{352} &  
\frac{137}{352}
 }  \right) \\ 
M_{3}[4,G_g,1] &=&  \left( \matrix{  
- \frac{545}{1392}   &  
-\frac{295\,{\sqrt{\frac{15}{11}}}}{928} &  
\frac{75}{32\,{\sqrt{319}}} \cr 
-\frac{295\,{\sqrt{\frac{15}{11}}}}{928} &  
- \frac{25291}{61248}   &  
-\frac{405\,{\sqrt{\frac{15}{29}}}}{704} \cr 
\frac{75}{32\,{\sqrt{319}}} &  
-\frac{405\,{\sqrt{\frac{15}{29}}}}{704} &  
\frac{115}{2112}
 }  \right) \\ 
M_{4}[4,G_g,1] &=&  \left( \matrix{  
- \frac{31}{464}   &  
\frac{5\,{\sqrt{\frac{15}{11}}}}{928} &  
-\frac{499}{96\,{\sqrt{319}}} \cr 
\frac{5\,{\sqrt{\frac{15}{11}}}}{928} &  
\frac{65833}{61248} &  
-\frac{1435\,{\sqrt{\frac{5}{87}}}}{704} \cr 
-\frac{499}{96\,{\sqrt{319}}} &  
-\frac{1435\,{\sqrt{\frac{5}{87}}}}{704} &  
\frac{53}{704}
 }  \right) \\ 
M_{5}[4,G_g,1] &=&  \left( \matrix{  
-\frac{55\,{\sqrt{5}}}{232} &  
\frac{2251}{464\,{\sqrt{33}}} &  
-\frac{65\,{\sqrt{\frac{5}{319}}}}{16} \cr 
\frac{2251}{464\,{\sqrt{33}}} &  
-\frac{2945\,{\sqrt{5}}}{10208} &  
-\frac{883}{352\,{\sqrt{87}}} \cr 
-\frac{65\,{\sqrt{\frac{5}{319}}}}{16} &  
-\frac{883}{352\,{\sqrt{87}}} &  
\frac{185\,{\sqrt{5}}}{352}
 }  \right) \\ 
\end{eqnarray*}
\begin{eqnarray*}
M_{2}[4,H_g,1] &=&  \left( \matrix{  
- \frac{2}{7}   &  
\frac{{\sqrt{3}}}{7} &  
-\frac{2}{3\,{\sqrt{7}}} \cr 
\frac{{\sqrt{3}}}{7} &  
\frac{53}{210} &  
-\frac{2}{5\,{\sqrt{21}}} \cr 
-\frac{2}{3\,{\sqrt{7}}} &  
-\frac{2}{5\,{\sqrt{21}}} &  
- \frac{3}{10}  
 }  \right) \\ 
M_{3}[4,H_g,1] &=&  \left( \matrix{  
- \frac{5}{42}   &  
-\frac{5\,{\sqrt{3}}}{14} &  
0 \cr 
-\frac{5\,{\sqrt{3}}}{14} &  
- \frac{25}{84}   &  
0 \cr 
0 &  
0 &  
\frac{5}{12}
 }  \right) \\ 
M_{4}[4,H_g,1] &=&  \left( \matrix{  
\frac{1}{14} &  
\frac{3\,{\sqrt{3}}}{14} &  
\frac{2}{3\,{\sqrt{7}}} \cr 
\frac{3\,{\sqrt{3}}}{14} &  
- \frac{121}{420}   &  
\frac{2}{5\,{\sqrt{21}}} \cr 
\frac{2}{3\,{\sqrt{7}}} &  
\frac{2}{5\,{\sqrt{21}}} &  
- \frac{9}{20}  
 }  \right) \\ 
M_{5}[4,H_g,1] &=&  \left( \matrix{  
\frac{2\,{\sqrt{5}}}{7} &  
-\frac{1}{7\,{\sqrt{15}}} &  
-\frac{2}{{\sqrt{35}}} \cr 
-\frac{1}{7\,{\sqrt{15}}} &  
-\frac{13}{14\,{\sqrt{5}}} &  
-\frac{4}{{\sqrt{105}}} \cr 
-\frac{2}{{\sqrt{35}}} &  
-\frac{4}{{\sqrt{105}}} &  
-\frac{1}{2\,{\sqrt{5}}}
 }  \right) \\ 
\end{eqnarray*}
\begin{eqnarray*}
M_{2}[5,G_u,1/2] &=&  \left( \matrix{  
\frac{281}{576} &  
\frac{{\sqrt{10}}}{21} &  
-\frac{775}{448\,{\sqrt{1983}}} &  
-\frac{5}{12\,{\sqrt{17186}}} &  
\frac{5\,{\sqrt{\frac{5}{78}}}}{8} \cr 
\frac{{\sqrt{10}}}{21} &  
\frac{25}{392} &  
-\frac{2465\,{\sqrt{\frac{5}{3966}}}}{147} &  
-\frac{575\,{\sqrt{\frac{5}{8593}}}}{63} &  
-\frac{5}{168\,{\sqrt{39}}} \cr 
-\frac{775}{448\,{\sqrt{1983}}} &  
-\frac{2465\,{\sqrt{\frac{5}{3966}}}}{147} &  
\frac{497123}{18656064} &  
-\frac{96685}{55524\,{\sqrt{78}}} &  
\frac{901\,{\sqrt{\frac{5}{17186}}}}{168} \cr 
-\frac{5}{12\,{\sqrt{17186}}} &  
-\frac{575\,{\sqrt{\frac{5}{8593}}}}{63} &  
-\frac{96685}{55524\,{\sqrt{78}}} &  
\frac{2953}{77337} &  
\frac{428\,{\sqrt{\frac{5}{1983}}}}{39} \cr 
\frac{5\,{\sqrt{\frac{5}{78}}}}{8} &  
-\frac{5}{168\,{\sqrt{39}}} &  
\frac{901\,{\sqrt{\frac{5}{17186}}}}{168} &  
\frac{428\,{\sqrt{\frac{5}{1983}}}}{39} &  
\frac{151}{936}
 }  \right) \\ 
M_{3}[5,G_u,1/2] &=&  \left( \matrix{  
- \frac{421}{1152}   &  
-\frac{11\,{\sqrt{\frac{5}{2}}}}{56} &  
-\frac{5335\,{\sqrt{\frac{3}{661}}}}{896} &  
-\frac{205}{8\,{\sqrt{17186}}} &  
-\frac{3\,{\sqrt{\frac{15}{26}}}}{16} \cr 
-\frac{11\,{\sqrt{\frac{5}{2}}}}{56} &  
\frac{2635}{7056} &  
\frac{4105\,{\sqrt{\frac{5}{3966}}}}{392} &  
-\frac{2005\,{\sqrt{\frac{5}{8593}}}}{84} &  
-\frac{55\,{\sqrt{\frac{3}{13}}}}{112} \cr 
-\frac{5335\,{\sqrt{\frac{3}{661}}}}{896} &  
\frac{4105\,{\sqrt{\frac{5}{3966}}}}{392} &  
\frac{25509209}{37312128} &  
\frac{78395}{37016\,{\sqrt{78}}} &  
\frac{333\,{\sqrt{\frac{5}{17186}}}}{112} \cr 
-\frac{205}{8\,{\sqrt{17186}}} &  
-\frac{2005\,{\sqrt{\frac{5}{8593}}}}{84} &  
\frac{78395}{37016\,{\sqrt{78}}} &  
\frac{8767}{77337} &  
-\frac{199\,{\sqrt{\frac{15}{661}}}}{52} \cr 
-\frac{3\,{\sqrt{\frac{15}{26}}}}{16} &  
-\frac{55\,{\sqrt{\frac{3}{13}}}}{112} &  
\frac{333\,{\sqrt{\frac{5}{17186}}}}{112} &  
-\frac{199\,{\sqrt{\frac{15}{661}}}}{52} &  
\frac{1405}{1872}
 }  \right) \\ 
M_{4}[5,G_u,1/2] &=&  \left( \matrix{  
\frac{59}{1152} &  
\frac{17\,{\sqrt{\frac{5}{2}}}}{168} &  
\frac{11675}{896\,{\sqrt{1983}}} &  
\frac{55\,{\sqrt{\frac{13}{1322}}}}{24} &  
-\frac{{\sqrt{\frac{65}{6}}}}{16} \cr 
\frac{17\,{\sqrt{\frac{5}{2}}}}{168} &  
\frac{195}{784} &  
\frac{4285\,{\sqrt{\frac{5}{3966}}}}{1176} &  
\frac{3245\,{\sqrt{\frac{5}{8593}}}}{252} &  
\frac{505}{336\,{\sqrt{39}}} \cr 
\frac{11675}{896\,{\sqrt{1983}}} &  
\frac{4285\,{\sqrt{\frac{5}{3966}}}}{1176} &  
- \frac{8816263}{37312128}   &  
-\frac{2305\,{\sqrt{\frac{13}{6}}}}{111048} &  
\frac{463\,{\sqrt{\frac{65}{1322}}}}{336} \cr 
\frac{55\,{\sqrt{\frac{13}{1322}}}}{24} &  
\frac{3245\,{\sqrt{\frac{5}{8593}}}}{252} &  
-\frac{2305\,{\sqrt{\frac{13}{6}}}}{111048} &  
\frac{29573}{77337} &  
-\frac{191\,{\sqrt{\frac{5}{1983}}}}{156} \cr 
-\frac{{\sqrt{\frac{65}{6}}}}{16} &  
\frac{505}{336\,{\sqrt{39}}} &  
\frac{463\,{\sqrt{\frac{65}{1322}}}}{336} &  
-\frac{191\,{\sqrt{\frac{5}{1983}}}}{156} &  
- \frac{419}{1872}  
 }  \right) \\ 
M_{5}[5,G_u,1/2] &=&  \left( \matrix{  
-\frac{19\,{\sqrt{5}}}{64} &  
\frac{1}{42\,{\sqrt{2}}} &  
-\frac{2795\,{\sqrt{\frac{5}{1983}}}}{448} &  
-\frac{41\,{\sqrt{\frac{5}{17186}}}}{6} &  
-\frac{15\,{\sqrt{\frac{3}{26}}}}{8} \cr 
\frac{1}{42\,{\sqrt{2}}} &  
-\frac{529\,{\sqrt{5}}}{1176} &  
-\frac{465\,{\sqrt{\frac{3}{1322}}}}{98} &  
\frac{4555}{42\,{\sqrt{8593}}} &  
-\frac{23\,{\sqrt{\frac{5}{39}}}}{56} \cr 
-\frac{2795\,{\sqrt{\frac{5}{1983}}}}{448} &  
-\frac{465\,{\sqrt{\frac{3}{1322}}}}{98} &  
\frac{917375\,{\sqrt{5}}}{2072896} &  
-\frac{4561\,{\sqrt{\frac{15}{26}}}}{9254} &  
\frac{4945}{56\,{\sqrt{17186}}} \cr 
-\frac{41\,{\sqrt{\frac{5}{17186}}}}{6} &  
\frac{4555}{42\,{\sqrt{8593}}} &  
-\frac{4561\,{\sqrt{\frac{15}{26}}}}{9254} &  
-\frac{6784\,{\sqrt{5}}}{25779} &  
\frac{341}{26\,{\sqrt{1983}}} \cr 
-\frac{15\,{\sqrt{\frac{3}{26}}}}{8} &  
-\frac{23\,{\sqrt{\frac{5}{39}}}}{56} &  
\frac{4945}{56\,{\sqrt{17186}}} &  
\frac{341}{26\,{\sqrt{1983}}} &  
\frac{59\,{\sqrt{5}}}{104}
 }  \right) \\ 
\end{eqnarray*}
\begin{eqnarray*}
M_{2}[5,H_u,1/2] &=&  \left( \matrix{  
\frac{8}{9} &  
-\frac{8}{3\,{\sqrt{21}}} &  
\frac{4\,{\sqrt{\frac{10}{21}}}}{3} &  
-\frac{{\sqrt{10}}}{9} &  
-\frac{4\,{\sqrt{\frac{2}{21}}}}{3} &  
\frac{4\,{\sqrt{\frac{10}{21}}}}{9} &  
\frac{5\,{\sqrt{\frac{2}{3}}}}{9} \cr 
-\frac{8}{3\,{\sqrt{21}}} &  
- \frac{8}{63}   &  
-\frac{10\,{\sqrt{10}}}{63} &  
\frac{{\sqrt{\frac{10}{21}}}}{9} &  
-\frac{8\,{\sqrt{2}}}{63} &  
-\frac{13\,{\sqrt{10}}}{189} &  
\frac{4\,{\sqrt{\frac{2}{7}}}}{27} \cr 
\frac{4\,{\sqrt{\frac{10}{21}}}}{3} &  
-\frac{10\,{\sqrt{10}}}{63} &  
\frac{22}{63} &  
-\frac{5}{9\,{\sqrt{21}}} &  
-\frac{5\,{\sqrt{5}}}{126} &  
- \frac{59}{378}   &  
-\frac{4\,{\sqrt{\frac{5}{7}}}}{27} \cr 
-\frac{{\sqrt{10}}}{9} &  
\frac{{\sqrt{\frac{10}{21}}}}{9} &  
-\frac{5}{9\,{\sqrt{21}}} &  
\frac{7}{54} &  
\frac{{\sqrt{\frac{35}{3}}}}{9} &  
-\frac{5\,{\sqrt{\frac{7}{3}}}}{27} &  
-\frac{2\,{\sqrt{\frac{5}{3}}}}{27} \cr 
-\frac{4\,{\sqrt{\frac{2}{21}}}}{3} &  
-\frac{8\,{\sqrt{2}}}{63} &  
-\frac{5\,{\sqrt{5}}}{126} &  
\frac{{\sqrt{\frac{35}{3}}}}{9} &  
0 &  
\frac{{\sqrt{5}}}{54} &  
-\frac{26}{27\,{\sqrt{7}}} \cr 
\frac{4\,{\sqrt{\frac{10}{21}}}}{9} &  
-\frac{13\,{\sqrt{10}}}{189} &  
- \frac{59}{378}   &  
-\frac{5\,{\sqrt{\frac{7}{3}}}}{27} &  
\frac{{\sqrt{5}}}{54} &  
\frac{2}{81} &  
\frac{26\,{\sqrt{\frac{5}{7}}}}{81} \cr 
\frac{5\,{\sqrt{\frac{2}{3}}}}{9} &  
\frac{4\,{\sqrt{\frac{2}{7}}}}{27} &  
-\frac{4\,{\sqrt{\frac{5}{7}}}}{27} &  
-\frac{2\,{\sqrt{\frac{5}{3}}}}{27} &  
-\frac{26}{27\,{\sqrt{7}}} &  
\frac{26\,{\sqrt{\frac{5}{7}}}}{81} &  
- \frac{7}{162}  
 }  \right) \\ 
M_{3}[5,H_u,1/2] &=&  \left( \matrix{  
\frac{10}{9} &  
0 &  
0 &  
0 &  
0 &  
0 &  
0 \cr 
0 &  
\frac{10}{63} &  
\frac{{\sqrt{10}}}{7} &  
0 &  
-\frac{5}{21\,{\sqrt{2}}} &  
-\frac{{\sqrt{\frac{5}{2}}}}{7} &  
\frac{1}{{\sqrt{14}}} \cr 
0 &  
\frac{{\sqrt{10}}}{7} &  
- \frac{17}{63}   &  
-\frac{{\sqrt{\frac{7}{3}}}}{12} &  
\frac{{\sqrt{5}}}{28} &  
- \frac{71}{252}   &  
-\frac{11\,{\sqrt{\frac{5}{7}}}}{36} \cr 
0 &  
0 &  
-\frac{{\sqrt{\frac{7}{3}}}}{12} &  
\frac{2}{9} &  
0 &  
-\frac{{\sqrt{\frac{7}{3}}}}{36} &  
\frac{5\,{\sqrt{\frac{5}{3}}}}{36} \cr 
0 &  
-\frac{5}{21\,{\sqrt{2}}} &  
\frac{{\sqrt{5}}}{28} &  
0 &  
\frac{5}{18} &  
\frac{{\sqrt{5}}}{4} &  
\frac{1}{2\,{\sqrt{7}}} \cr 
0 &  
-\frac{{\sqrt{\frac{5}{2}}}}{7} &  
- \frac{71}{252}   &  
-\frac{{\sqrt{\frac{7}{3}}}}{36} &  
\frac{{\sqrt{5}}}{4} &  
\frac{13}{27} &  
-\frac{11\,{\sqrt{\frac{5}{7}}}}{108} \cr 
0 &  
\frac{1}{{\sqrt{14}}} &  
-\frac{11\,{\sqrt{\frac{5}{7}}}}{36} &  
\frac{5\,{\sqrt{\frac{5}{3}}}}{36} &  
\frac{1}{2\,{\sqrt{7}}} &  
-\frac{11\,{\sqrt{\frac{5}{7}}}}{108} &  
- \frac{11}{54}  
 }  \right) \\ 
M_{4}[5,H_u,1/2] &=&  \left( \matrix{  
\frac{10}{9} &  
\frac{8}{3\,{\sqrt{21}}} &  
-\frac{4\,{\sqrt{\frac{10}{21}}}}{3} &  
\frac{{\sqrt{10}}}{9} &  
\frac{4\,{\sqrt{\frac{2}{21}}}}{3} &  
-\frac{4\,{\sqrt{\frac{10}{21}}}}{9} &  
-\frac{5\,{\sqrt{\frac{2}{3}}}}{9} \cr 
\frac{8}{3\,{\sqrt{21}}} &  
\frac{26}{63} &  
\frac{{\sqrt{10}}}{63} &  
-\frac{{\sqrt{\frac{10}{21}}}}{9} &  
-\frac{11}{63\,{\sqrt{2}}} &  
\frac{53\,{\sqrt{\frac{5}{2}}}}{189} &  
-\frac{5\,{\sqrt{\frac{7}{2}}}}{27} \cr 
-\frac{4\,{\sqrt{\frac{10}{21}}}}{3} &  
\frac{{\sqrt{10}}}{63} &  
\frac{23}{63} &  
\frac{83}{36\,{\sqrt{21}}} &  
\frac{{\sqrt{5}}}{252} &  
\frac{163}{756} &  
\frac{{\sqrt{35}}}{108} \cr 
\frac{{\sqrt{10}}}{9} &  
-\frac{{\sqrt{\frac{10}{21}}}}{9} &  
\frac{83}{36\,{\sqrt{21}}} &  
\frac{1}{27} &  
-\frac{{\sqrt{\frac{35}{3}}}}{9} &  
\frac{29\,{\sqrt{\frac{7}{3}}}}{108} &  
-\frac{37\,{\sqrt{\frac{5}{3}}}}{108} \cr 
\frac{4\,{\sqrt{\frac{2}{21}}}}{3} &  
-\frac{11}{63\,{\sqrt{2}}} &  
\frac{{\sqrt{5}}}{252} &  
-\frac{{\sqrt{\frac{35}{3}}}}{9} &  
\frac{1}{2} &  
-\frac{29\,{\sqrt{5}}}{108} &  
\frac{25}{54\,{\sqrt{7}}} \cr 
-\frac{4\,{\sqrt{\frac{10}{21}}}}{9} &  
\frac{53\,{\sqrt{\frac{5}{2}}}}{189} &  
\frac{163}{756} &  
\frac{29\,{\sqrt{\frac{7}{3}}}}{108} &  
-\frac{29\,{\sqrt{5}}}{108} &  
\frac{43}{81} &  
-\frac{113\,{\sqrt{\frac{5}{7}}}}{324} \cr 
-\frac{5\,{\sqrt{\frac{2}{3}}}}{9} &  
-\frac{5\,{\sqrt{\frac{7}{2}}}}{27} &  
\frac{{\sqrt{35}}}{108} &  
-\frac{37\,{\sqrt{\frac{5}{3}}}}{108} &  
\frac{25}{54\,{\sqrt{7}}} &  
-\frac{113\,{\sqrt{\frac{5}{7}}}}{324} &  
\frac{133}{162}
 }  \right) \\ 
M_{5}[5,H_u,1/2] &=&  \left( \matrix{  
0 &  
4\,{\sqrt{\frac{5}{21}}} &  
4\,{\sqrt{\frac{2}{21}}} &  
-\frac{{\sqrt{2}}}{3} &  
2\,{\sqrt{\frac{10}{21}}} &  
\frac{4\,{\sqrt{\frac{2}{21}}}}{3} &  
\frac{{\sqrt{\frac{10}{3}}}}{3} \cr 
4\,{\sqrt{\frac{5}{21}}} &  
\frac{8\,{\sqrt{5}}}{21} &  
-\frac{2\,{\sqrt{2}}}{7} &  
\frac{{\sqrt{\frac{2}{21}}}}{3} &  
\frac{{\sqrt{10}}}{21} &  
-\frac{19\,{\sqrt{2}}}{63} &  
\frac{2\,{\sqrt{\frac{10}{7}}}}{9} \cr 
4\,{\sqrt{\frac{2}{21}}} &  
-\frac{2\,{\sqrt{2}}}{7} &  
-\frac{8\,{\sqrt{5}}}{21} &  
\frac{5\,{\sqrt{\frac{5}{21}}}}{6} &  
- \frac{1}{14}   &  
\frac{31\,{\sqrt{5}}}{126} &  
-\frac{13}{18\,{\sqrt{7}}} \cr 
-\frac{{\sqrt{2}}}{3} &  
\frac{{\sqrt{\frac{2}{21}}}}{3} &  
\frac{5\,{\sqrt{\frac{5}{21}}}}{6} &  
0 &  
\frac{{\sqrt{\frac{7}{3}}}}{3} &  
\frac{{\sqrt{\frac{35}{3}}}}{6} &  
-\frac{1}{6\,{\sqrt{3}}} \cr 
2\,{\sqrt{\frac{10}{21}}} &  
\frac{{\sqrt{10}}}{21} &  
- \frac{1}{14}   &  
\frac{{\sqrt{\frac{7}{3}}}}{3} &  
\frac{{\sqrt{5}}}{3} &  
\frac{7}{18} &  
-\frac{4\,{\sqrt{\frac{5}{7}}}}{9} \cr 
\frac{4\,{\sqrt{\frac{2}{21}}}}{3} &  
-\frac{19\,{\sqrt{2}}}{63} &  
\frac{31\,{\sqrt{5}}}{126} &  
\frac{{\sqrt{\frac{35}{3}}}}{6} &  
\frac{7}{18} &  
-\frac{4\,{\sqrt{5}}}{9} &  
-\frac{31}{18\,{\sqrt{7}}} \cr 
\frac{{\sqrt{\frac{10}{3}}}}{3} &  
\frac{2\,{\sqrt{\frac{10}{7}}}}{9} &  
-\frac{13}{18\,{\sqrt{7}}} &  
-\frac{1}{6\,{\sqrt{3}}} &  
-\frac{4\,{\sqrt{\frac{5}{7}}}}{9} &  
-\frac{31}{18\,{\sqrt{7}}} &  
\frac{{\sqrt{5}}}{9}
 }  \right) \\ 
\end{eqnarray*}
 


We report here the explicit expressions for the quantities $\Delta_1 ...
\Delta_5$ of Table~\ref{analyticMultiplet}, originated by diagonalizations
of $2\times 2$ blocks:
\begin{eqnarray}
\Delta_1 &=& \frac{1}{6} \left[ 4 \, F_2^2 - 12 \, F_2 F_3 + 9 \, F_3^2 + 
4 \, F_2 F_4 - 6 \, F_3 F_4 + F_4^2 + 9 \, F_5^2 \right]^{1/2} \\
\Delta_2 &=& \frac{1}{12} \left[ 4 \, F_2^2 + 12 \, F_2 F_3 + 9 \, F_3^2 
- 20 \, F_2 F_4 - 30 \, F_3 F_4 + 25 \, F_4^2 + 300 \, F_5^2 \right]^{1/2} \\
\Delta_3 &=& \frac{1}{12} \left[ 4 \, F_2^2 + 36 \, F_2 F_3 + 81 \, F_3^2 
-44 \, F_2 F_4 - 198 \, F_3 F_4 + 121 \, F_4^2 + 288 \, F_5^2 \right]^{1/2}\\
\Delta_{4/5} &=& \frac{1}{24} \left[
964 \, F_2^2 - 324\, F_2 F_3 + 81 \, F_3^2 
-1604 \,F_2 F_4  + 162 \, F_3 F_4 + 721 \, F_4^2 \right.\\
 && \left.\quad\ 
\mp 984 \, \sqrt{5} \, F_2 F_5
\pm 108 \, \sqrt{5} \, F_3 F_5
\pm 876 \, \sqrt{5} \, F_4 F_5
 +1332 \, F_5^2 \right]^{1/2} .
\end{eqnarray}

\section*{References}



\begin{itemize}
\item 
Antropov,  V.\ P., Gunnarsson, O.,  and Jepsen, O., 1992, Phys.\ Rev.\ B {\bf 46}, 13647. 
\item 
Arovas,  D.\ P., and Auerbach A., 1995, Phys.\ Rev.\ B {\bf 52}, 10114. 
\item 
Baroni, S., Dal Corso, A., de Gironcoli, S., and Giannozzi, P., 2002,
http://www.pwscf.org 
\item 
Benning,  P.\ J., Stepniak, F.,  and Weaver, J.\ H., 1993, Phys.\ Rev.\ B {\bf 48}, 9086. 
\item 
Brouet,  V., Alloul, H., Quere, F., Baumgartner, G.,  and Forro, L., 1999, Phys.\ Rev.\ Lett.\ {\bf 82}, 2131. 
\item 
Brouet,  V., Alloul, H., Le, T.\ N., Garaj, S.,  and Forro, L., 2001, Phys.\ Rev.\ Lett.\ {\bf 86}, 4680. 
\item 
Butler,  P.\ H., 1981, {\it Point Group Symmetry Applications} (Ple\-num, New York). 
\item 
Capone,  M., Fabrizio, M., Giannozzi, P.,  and Tosatti, E., 2000, Phys.\ Rev.\ B {\bf 62}, 7619. 
\item 
Capone,  M., Fabrizio, M.,  and Tosatti, E., 2001, Phys.\ Rev.\ Lett.\ {\bf 86}, 5361. 
\item 
Capone,  M., Fabrizio, M., Castellani, C.,  and Tosatti, E., 2002, Science {\bf 296}, 2364. 
\item 
Chang,  A.\ H.\ H., Ermler, W.\ C.,  and Pitzer, R.\ M., 1991, J.\ Phys.\ Chem.\ {\bf 95}, 9288. 
\item 
Cowan,  R.\ D., 1981, {\it The Theory of Atomic Structure, Spectra} (Univ.\ of California Press, Berkeley-CA). 
\item 
Dederichs,  P.\ H., Bl{\"u}gel, S., Zeller, R.,  and Akai, H., 1984, Phys.\ Rev. Lett, {\bf 53}, 2512. 
\item 
Fabrizio,  M.,  and Tosatti, E., 1997, Phys.\ Rev.\ B {\bf 55}, 13465. 
\item 
Favot, F.,  and Dal Corso, A., 1999, Phys.\ Rev.\ B {\bf 60}, 11427. 
\item 
Gunnarsson,  O., Andersen, O.\ K., Jepsen, O.,  and Zaanen, J., 1989, Phys.\ Rev.\ B {\bf 39}, 1708. 
\item 
Gunnarsson,  O., Handschuh, H., Bechthold, P.\ S., Kessler, B., Gantef\"or, G.,  and Eberhardt, W., 1995, Phys.\ Rev.\ Lett.\ {\bf 74}, 1875;  and Gunnarsson, O., 1995, Phys.\ Rev.\ B {\bf 51}, 3493 (1995). 
\item 
Gunnarsson,  O., 1997, Rev.\ Mod.\ Phys.\ {\bf 69}, 575. 
\item 
Han,  J.\ E.,  and Gunnarsson, O., 2000, Physica B {\bf 292}, 196. 
\item 
Janak,  J.\ F., 1978, Phys.\ Rev.\ B {\bf 18}, 7165. 
\item 
Jarvis,  M.\ R., White, I.\ D., Godby, R.\ W.,  and Payne, M.\ C., 1997, Phys.\ Rev.\ B {\bf 56}, 14972. 
\item 
Jones,  R.\ O.,  and Gunnarsson, O., 1989, Rev.\ Mod.\ Phys.\ {\bf 61}, 689. 
\item 
Kiefl,  R.\ F., Duty, T.\ L., Schneider, J.\ W., MacFarlane, A., Chow, K., Elzey, J.\ W., Mendels, P., Morris, G.\ D., Brewer, J.\ H., Ansaldo, E.\ J., Niedermayer, C., Noakes, D.\ R., Stronach, C.\ E., Hitti, B.,  and Fischer, J.\ E., 1992, Phys.\ Rev.\ Lett.\ {\bf 69}, 2005. 
\item 
Lo,  E.,  and Judd, B.\ R., 1999, Phys.\ Rev.\ Lett.\ {\bf 82}, 3224. 
\item 
Lof,  R.\ W., van Veenendaal, M.\ A., Koopmans, B., Jonkman, H.\ T.,  and Sawatzky, G.\ A., 1992, Phys.\ Rev.\ Lett.\ {\bf 68}, 3924. 
\item 
Lukyanchuk,  I., Kirova, N., Rachdi, F., Goze, C., Molinie, P.,  and Mehring, M., 1995, Phys.\ Rev.\ B {\bf 51}, 3978. 
\item 
Makov,  G.,  and Payne, M.\ C., 1995, Phys.\ Rev.\ B {\bf 51}, 4014. 
\item 
Manini,  N., Tosatti, E.,  and Auerbach, A., 1994, Phys.\ Rev.\ B {\bf 49}, 13008. 
\item 
Manini, N.,  and Tosatti, E., 1998, Phys.\ Rev.\ B {\bf 58}, 782. 
\item 
Manini,  N.,  and De Los Rios, P., 2000, Phys.\ Rev.\ B {\bf 62}, 29. 
\item 
Manini,  N., Dal Corso, A., Fabrizio, M.,  and Tosatti, E., 2001, Phil.\ Mag.\ B {\bf 81}, 793. 
\item 
Martin,  R.\ L.,  and Ritchie, J.\ P., 1993, Phys.\ Rev.\ B {\bf 48}, 4845. 
\item 
Nikolaev,  A.\ V.,  and Michel, K.\ H., 2002, 
J. Chem. Phys. {\bf 117}, 4761 (2002).
\item 
O'Brien,  M.\ C.\ M., 1972, J.\ Phys.\ C {\bf 5}, 2045. 
\item 
O'Brien,  M.\ C.\ M., 1996, Phys.\ Rev.\ B {\bf 53}, 3775. 
\item 
Oliva,  J.\ M., 1997, Phys.\ Lett.\ A {\bf 234}, 41. 
\item 
Perdew, J., 1985, in {\it Density Functional Methods in Physics}, edited by
Dreizler, \ R.\ M., da Providencia, J.\ (Plenum, New York, Series
B, Vol. 123, p. 265).
\item 
Petersilka,  M., Gossmann, U.\ J.,  and Gross, E.\ K.\ U., 2000, Phys.\ Rev.\ Lett.\ {\bf 76}, 1212; Petersilka, M.,  and Gross, E.\ K.\ U., 2000, Int.\ J.\ Quant.\ Chem.\ Symp.\ {\bf 30}, 1393 (1996); Grabo, T., Petersilka, M.,  and Gross, E.\ K.\ U., 2000, Journal of Molecular Structure (Theochem) {\bf 501}, 353 (2000). 
\item 
Plakhutin,  B.\ N.,  and Carb{\'o}-Dorca, R., 2000, Phys.\ Lett.\ A {\bf 267}, 370. 
\item 
Prassides,  K., Margadonna, S., Arcon, D., Lappas, A., Shimoda, H.,  and Iwasa, Y., 1999, J.\ Am.\ Chem.\ Soc.\ {\bf 121}, 11227. 
\item 
Ramirez,  A.\ P., 1994, Supercond.\ Review {\bf 1}, 1. 
\item 
Schilder,  A., Klos, H., Rystau, I.,  and Sch\"utz, W., B.\ Gotschy, 1994, Phys.\ Rev. Lett.\ {\bf 73}, 1299. 
\item 
Schipper,  P.\ R.\ T., Gritsenko, O.\ V.,  and Baerends, E.\ J., 1998, Theor.\ Chem.\ Acc.\ {\bf 99}, 329. 
\item 
Sch\"on,  J.\ H., Kloc, Ch.,  and Batlogg, B., 2000a, Nature {\bf 408}, 549. 
\item 
Sch\"on,  J.\ H., Kloc, Ch., Haddon, R.\ C.,  and Batlogg, B., 2000b, Science {\bf 288}, 656. 
\item 
Tosatti,  E., Manini, N.,  and Gunnarsson, O., 1996, Phys.\ Rev.\ B {\bf 54}, 17184. 
\item 
Vanderbilt,  D., 1990, Phys.\ Rev.\ B {\bf 41}, 7892. 
\item 
Zimmer,  G., Mehring, M., Goze, C.,  and Rachdi, F., 1995, in {\it Physics,
Chemistry of Fullerenes, Derivatives}, edited by Kuzmany, H., Fink, J.,
Mehring, M.,  and Roth, S. (World Scientific, Singapore), p.\ 452.

\end{itemize}


\begin{table}[ht]
\begin{center}
\begin{tabular}{cll}
\hline
\hline
State  & $c_i$ [eV] & $c_i$ [eV] \\
$|i\rangle$ & (from $E^{\rm tot}$) & (from $\epsilon_i$) \\
\hline
$|\uparrow\downarrow,\uparrow\downarrow,\uparrow\downarrow\rangle$ & 3.06855 & 3.06850 \\
$|\uparrow,\uparrow,\uparrow\rangle$ & 3.04652 & 3.04659 \\
$|0,\uparrow,0\rangle$ & 3.06581 & 3.06650 \\
\hline
\hline
\end{tabular}
\end{center}
\caption{
Quadratic coefficients $c_i$ for the different configurations used in the
C$_{60}^{n-}$ calculations, obtained from the total-energy curvature and
from the linear dependence of the single-particle KS energies.
The Slater determinant states $|i\rangle$ are specified by the individual
occupancy of the LUMO orbitals labeled by $m=-1,0,1$.
\label{c60-coefs:tab}
}
\end{table}

\begin{table}[ht]
\begin{center}
\begin{tabular}{cll}
\hline
\hline
State  & $c_i$ [eV] & $c_i$ [eV] \\
$|i\rangle$ & (from $E^{\rm tot}$) & (from $\epsilon_i$) \\
\hline
$\ket{0,0,\up \down,0,0}          $ & $3.20874$ & $3.20251$ \\ 
$\ket{0,\up \down, 0,\up \down,0} $ & $3.12273$ & $3.12234$\\
$\ket{\up \down,0,0,0,\up \down}  $ & $3.12349$ & $3.12274$\\
$\ket{\up \down,\up \down,\up \down,\up \down,\up \down}$ & $3.10285$ & $3.10090$ \\
\hline
$\ket{\up,\up,\up,\up,\up}$ & $3.08025$ & $3.07928$ \\
$\ket{0,\up,\up,\up,0}    $ & $3.07822$ & $3.08319$ \\
$\ket{\up,0, \up,0, \up}  $ & $3.08654$ & $3.08415$ \\
$\ket{0,0,\up,0,0}        $ & $3.15757$ & $3.15165$ \\
\hline
\hline
\end{tabular}
\end{center}
\caption{
Quadratic coefficients $c_i$ for the different configurations used in the
C$_{60}^{n+}$ calculations, obtained with the two methods.
The $|n_{-2},n_{-1},n_{0},n_{1},n_{2}\rangle$ notation specifies
the occupancies of the $m=-2,-1,0,1,2$ HOMO orbitals in each Slater
determinant $i$.
\label{c60+coefs:tab}
}
\end{table}

\begin{table}[ht]
\begin{center}
\renewcommand{\arraystretch}{1.3}
\begin{tabular}{lclc}
\hline
\hline
State $|i\rangle$ &  $n_i$ &
\multicolumn{2}{c}{ 
$2 n_i^{-2}\left(\langle i|\hat{H}_{\rm e-e}|i\rangle + U n_i /2\right)$ 
} \\
& & Model: $ U + F[\ket{i}] $ & 
from Eq.~(\ref{parabola}):
$c_i + 2 \mu / n_i$
\\
\hline
$\ket{\up \down, \up \down, \up \down}$ & 6 &
$ U $ & 
$c_{\ket{\up \down, \up \down, \up \down}}+\frac 13 \mu$\\
$\ket{\up, \up, \up}$ & 3 &
$ U - \frac 23 J$ & 
$c_{\ket{\up, \up, \up}}+\frac 23 \mu$\\
$\ket{0,\up,0}$ & 1 & $U$ &
$c_{\ket{0,\up,0}}+ 2 \mu$\\
\hline
\hline
\end{tabular}
\renewcommand{\arraystretch}{1}
\caption{
Comparison of the energy expectation values in the model
(\ref{modelhamiltonian}) and from the LSDA extrapolation (\ref{parabola})
for the electron states considered in the calculation.
\label{comparison-:table}
}
\end{center}
\end{table}






\begin{table}[ht]
\begin{center}
\begin{tabular}{crrr}
\hline
\hline
Parameter & ($\hat{H}_{\rm e-e}$) &
($\hat{H}_{\rm e-vib}$, 2$^{\rm nd}$ order) & Total: (e-e)+(e-vib) \\
        &       [meV]   &       [meV]   &       [meV]   \\
\hline
$U$     & 3069          &         32    &       3101    \\
$J$     &   32          &        -57    &        -25    \\
\hline
\hline
\end{tabular}
\caption{
The Coulomb parameters for C$_{60}^{n-}$, as obtained from the LSDA
calculations, the effective parameters obtained from second-order treatment
of $\hat{H}_{\rm e-vib}$ [based on the couplings of
(Manini {\it et al.}\ 2001)], and the sum of the two contributions.
\label{parameters-:table}
}
\end{center}
\end{table}

\begin{table}[ht]
\begin{center}
\renewcommand{\arraystretch}{1.3}
\begin{tabular}{lclc}
\hline
\hline
State $|i\rangle$ &  $n_i$ &
\multicolumn{2}{c}{ 
$2 n_i^{-2}\left(\langle i|\hat{H}_{\rm e-e}|i\rangle + U n_i /2\right)$ 
} \\
& & Model: $ U + F[\ket{i}] $ & 
from Eq.~(\ref{parabola}):
$c_i + 2 \mu / n_i$
\\
\hline
$\ket{0, 0,\up \down, 0, 0}$ & 2 & 
$ U + \frac{2}{45} F_2 + \frac{41}{90} F_3 + \frac{1}{18} F_4 $ &
$c_{\ket{0, 0,\up \down, 0, 0}} + \mu$ \\
$\ket{0, \up \down,0,\up \down,0}$ & 4 &
$ U - \frac{1}{15} F_2 + \frac{11}{240} F_3 + \frac{11}{48} F_4
- \frac{\sqrt{5}}{8} F_5$ &
$c_{\ket{0, \up \down,0,\up \down,0}}+ \frac12 \mu$ \\
$\ket{\up \down, 0, 0, 0, \up \down}$ & 4 &
$ U - \frac{1}{15} F_2 + \frac{11}{240} F_3 + \frac{11}{48} F_4
+ \frac{\sqrt{5}}{8} F_5$ &
$c_{\ket{\up \down, 0, 0, 0, \up \down}}+ \frac 12 \mu$\\
$\ket{\up \down, \up \down, \up \down, \up \down, \up \down}$ & 10 &
$ U $ & 
$c_{\ket{\up \down, \up \down, \up \down, \up \down, \up \down}}+\frac 15 \mu$\\
\hline
$\ket{\up, \up, \up, \up, \up}$ & 5 &
$ U - \frac{4}{45} F_2 - \frac{1}{9} F_3 
- \frac{1}{9} F_4$ & 
$c_{\ket{\up, \up, \up, \up, \up}}+\frac 25 \mu$\\
$\ket{0, \up, \up, \up, 0}$ & 3 & 
$ U - \frac{4}{27} F_2 - \frac{5}{54} F_3 - \frac{1}{54} F_4 +
\frac{1}{3 \sqrt{5}} F_5 $ & 
$c_{\ket{0, \up, \up, \up, 0}}+\frac{2}{3} \mu$\\
$\ket{\up, 0, \up, 0, \up}$ & 3 & 
$ U - \frac{4}{27} F_2 - \frac{5}{54} F_3 - \frac{1}{54} F_4 -
\frac{1}{3 \sqrt{5}} F_5 $ & 
$c_{\ket{\up, 0, \up, 0, \up}}+\frac{2}{3} \mu$\\
$\ket{0,0,\up,0,0}$ & 1 & $U$ &
$c_{\ket{0,0,\up,0,0}}+ 2 \mu$\\
\hline
\hline
\end{tabular}
\renewcommand{\arraystretch}{1}
\caption{
Comparison of the  energy expectation values in the model
(\ref{modelhamiltonian}) and from the LSDA extrapolation (\ref{parabola})
for the hole states considered.
\label{comparison+:table}
}
\end{center}
\end{table}


 




%
%
%

\begin{table}[ht]
\begin{center}
\begin{tabular}{crrr}
\hline
\hline
Parameter & ($\hat{H}_{\rm e-e}$) &
($\hat{H}_{\rm e-vib}$, 2$^{\rm nd}$ order) & Total: (e-e)+(e-vib) \\
        &       [meV]   &       [meV]   &       [meV] \\
\hline
$F_1$   &15646 $\pm$  9 &        -18    &      15628    \\
$F_2$   &  105 $\pm$ 10 &        -62    &         42    \\
$F_3$   &  155 $\pm$  4 &       -173    &        -18    \\
$F_4$   &   47 $\pm$  5 &        -50    &         -3    \\
$F_5$   &    0 $\pm$  3 &        -14    &        -14    \\
$\mu$   &  -27 $\pm$  1 &               & \\
\hline
$U$     & 3097 $\pm$  1 &         27    &       3124    \\
$J$     &   60 $\pm$  1 &        -57    &          3    \\
\hline
\hline
\end{tabular}
\caption{
The Coulomb parameters for C$_{60}^{n+}$, as obtained from the LSDA
calculations described in the text.
Two of the tabulated parameters (e.g.\ $F_1$ and $J$) are linear
combinations of the five others.
The table also includes the effective (negative, anti-Hund) parameters
obtained from second-order treatment of $\hat{H}_{\rm e-vib}$ [based on the
electron-phonon couplings of Manini {\it et al.}\ (2001)],
and the sum of the two contributions.
\label{parameters+:table}
}
\end{center}
\end{table}

\begin{table}[ht]
\begin{center}
\begin{tabular}{c|rr}
\hline \hline
$n$  & $U^{\rm min}$ & $\Delta_{\rm spin}^{e-e}$  \\
\hline
1 & 3038 &      \\
2 & 3077 &  76  \\
3 & 3076 &  99  \\
4 & 3038 & 132  \\
5 & 3415 & 202  \\
\hline
\hline
\end{tabular}
\end{center}
\caption{
The Coulomb Hubbard $ U^{\rm min}
= E^{\rm min}(n+1) + E^{\rm min}(n-1) - 2 E^{\rm min}(n) $ and the
Coulomb spin-gap $\Delta_{\rm spin}^{e-e}$ to the first low-spin state.
\label{Umintable}
}
\end{table}

\begin{table}[ht]
\begin{center}
\renewcommand{\arraystretch}{1.2}
\[
\begin{array}{lcl}
\hline
\hline
\mbox{State}    & \mbox{spin}   & \mbox{Energy} \\
\mbox{symmetry}& S      &  E_{\rm mult} \\
\hline
n=2 & & \\
A_g    & 0   &  \frac{8}{9}  F_2.+ \frac{10}{9}  F_3 + \frac{10}{9}  F_4  \\
G_g    & 0   &  \frac{1}{18} F_2 - \frac{5}{36}  F_3 + \frac{25}{36} F_4 \\
H_g    & 0   &  \frac{2}{9}  F_2 + \frac{13}{36} F_3 + \frac{1}{36}  F_4 - \Delta_1\\
H_g    & 0   &  \frac{2}{9}  F_2 + \frac{13}{36} F_3 + \frac{1}{36}  F_4 + \Delta_1\\
T_{1g} & 1   & -\frac{4}{9}  F_2 - \frac{5}{36}  F_3 + \frac{7}{36}  F_4 + \frac{\sqrt{5}}{2} F_5 \\
T_{2g} & 1   & -\frac{4}{9}  F_2 - \frac{5}{36}  F_3 + \frac{7}{36}  F_4 - \frac{\sqrt{5}}{2} F_5 \\
G_g    & 1   &  \frac{7}{18} F_2 - \frac{5}{36}  F_3 - \frac{23}{36} F_4\\
\hline
n=3 & & \\
T_{1u} & 1/2 & -\frac{1}{6}  F_2 - \frac{5}{12} F_3 + \frac{11}{12} F_4 - \frac{\sqrt{5}}{2} F_5 \\
T_{1u} & 1/2 &  \frac{1}{3}  F_2 + \frac{1}{3}  F_3 - \frac{1}{3}   F_4 \\
T_{2u} & 1/2 &  \frac{1}{3}  F_2 + \frac{1}{3}  F_3 - \frac{1}{3}   F_4 \\
T_{2u} & 1/2 & -\frac{1}{6}  F_2 - \frac{5}{12} F_3 + \frac{11}{12} F_4 + \frac{\sqrt{5}}{2} F_5 \\
G_u    & 1/2 & -\frac{1}{3}  F_2 + \frac{7}{12} F_3 + \frac{1}{12}  F_4 - \Delta_2\\
G_u    & 1/2 & -\frac{1}{3}  F_2 + \frac{7}{12} F_3 + \frac{1}{12}  F_4 + \Delta_2\\
H_u\ [\times 4]    & 1/2 & \mbox{Eigenvalues}( H_{\rm mult}[3,H_u,1/2] ) \\
T_{1u} & 3/2 & -\frac{2}{3}  F_2 - \frac{5}{12} F_3 - \frac{1}{12}  F_4 + \frac{\sqrt{5}}{2} F_5 \\
T_{2u} & 3/2 & -\frac{2}{3}  F_2 - \frac{5}{12} F_3 - \frac{1}{12}  F_4 - \frac{\sqrt{5}}{2} F_5 \\
G_u    & 3/2 &  \frac{1}{6}  F_2 - \frac{5}{12} F_3 - \frac{11}{12} F_4\\
\hline
\end{array}
\begin{array}{lcl}
\hline
\hline
\mbox{State}    & \mbox{spin}   & \mbox{Energy} \\
\mbox{symmetry} & S     &  E_{\rm mult} \\
\hline
n=4 & & \\
A_g\    [\times 3] & 0 & \mbox{Eigenvalues}( H_{\rm mult}[4,A_g,0] ) \\
T_{1g} &  0  &  \frac{1}{2} F_2 - \frac{1}{12} F_3 + \frac{1}{4} F_4 - \frac{\sqrt{5}}{2} F_5  \\
T_{2g} &  0  &  \frac{1}{2} F_2 - \frac{1}{12} F_3 + \frac{1}{4} F_4 + \frac{\sqrt{5}}{2} F_5  \\
G_g\    [\times 4] & 0 & \mbox{Eigenvalues}( H_{\rm mult}[4,G_g,0] ) \\
H_g\    [\times 5] & 0 & \mbox{Eigenvalues}( H_{\rm mult}[4,H_g,0] ) \\
T_{1g}\ [\times 3] & 1 & \mbox{Eigenvalues}( H_{\rm mult}[4,T_{1g},1] ) \\
T_{2g}\ [\times 3] & 1 & \mbox{Eigenvalues}( H_{\rm mult}[4,T_{2g},1] ) \\
G_g\    [\times 3] & 1 & \mbox{Eigenvalues}( H_{\rm mult}[4,G_g,1] ) \\
H_g\    [\times 3] & 1 & \mbox{Eigenvalues}( H_{\rm mult}[4,H_g,1] ) \\
H_g    &  2  & -\frac{2}{3} F_2 - \frac{5}{6} F_3 - \frac{5}{6} F_4\\
\hline
n = 5 & & \\
A_u     & 1/2 & -\frac5{18} F_2 +\frac {13}{36} F_3 + \frac1{36}F_4 -\Delta_3\\
A_u     & 1/2 & -\frac5{18} F_2 +\frac {13}{36} F_3 + \frac1{36}F_4 +\Delta_3\\
T_{1u}  & 1/2 & \frac1{18} F_2 +\frac {13}{36} F_3 - \frac{11}{36}F_4 - \frac{\sqrt{5}}{2} F_5 \\
T_{1u}  & 1/2 & \frac{23}{36} F_2 +\frac {17}{72} F_3 + \frac{17}{72}F_4 + \frac{3 \sqrt{5}}4 F_5 -\Delta_4\\
T_{1u}  & 1/2 & \frac{23}{36} F_2 +\frac {17}{72} F_3 + \frac{17}{72}F_4 + \frac{3 \sqrt{5}}4 F_5 +\Delta_4\\
T_{2u}  & 1/2 & \frac1{18} F_2 +\frac {13}{36} F_3 - \frac{11}{36}F_4 + \frac{\sqrt{5}}{2} F_5 \\
T_{2u}  & 1/2 & \frac{23}{36} F_2 +\frac {17}{72} F_3 + \frac{17}{72}F_4 - \frac{3 \sqrt{5}}4 F_5 -\Delta_5\\
T_{2u}  & 1/2 & \frac{23}{36} F_2 +\frac {17}{72} F_3 + \frac{17}{72}F_4 - \frac{3 \sqrt{5}}4 F_5 +\Delta_5\\
G_u\    [\times 5] & 1/2 & \mbox{Eigenvalues}( H_{\rm mult}[5,G_u,1/2] ) \\
H_u\    [\times 7] & 1/2 & \mbox{Eigenvalues}( H_{\rm mult}[5,H_u,1/2] ) \\
T_{1u} & 3/2 &  \frac{2}{9}  F_2 - \frac{5}{36} F_3 - \frac{17}{36} F_4 - \frac{\sqrt{5}}{2} F_5 \\
T_{2u} & 3/2 &  \frac{2}{9}  F_2 - \frac{5}{36} F_3 - \frac{17}{36} F_4 + \frac{\sqrt{5}}{2} F_5 \\
G_u    & 3/2 &  -\frac{5}{18} F_2 - \frac{5}{36} F_3 - \frac{35}{36} F_4 \\
G_u    & 3/2 & -\frac{11}{18} F_2 - \frac{5}{36} F_3 + \frac{13}{36} F_4 \\
H_u    & 3/2 & -\frac{4}{9}  F_2 - \frac{23}{36} F_3 - \frac{11}{36} F_4 - \Delta_1 \\
H_u    & 3/2 & -\frac{4}{9}  F_2 - \frac{23}{36} F_3 - \frac{11}{36} F_4 + \Delta_1 \\
A_u    & 5/2 & -\frac{10}{9} F_2 - \frac{25}{18} F_3 - \frac{25}{18} F_4 \\
\hline
\hline
\end{array}
\]
\renewcommand{\arraystretch}{1.0}
\caption{
The Coulomb multiplets for C$_{60}^{n+}$ as a function of the e-e
parameters.
The model Hamiltonian (\ref{modelhamiltonian}) obeys particle-hole
symmetry: therefore the multiplet energies for $n>5$ holes equal those for
$(10-n)$ holes.
The non particle-hole symmetric contribution $\left[\epsilon n + U
n(n-1)/2\right]$ is left out in this table.
The $ H_{\rm mult}[n,\Lambda,S]$ and $\Delta_i$ quantities are defined in
Appendix~\ref{matrices}.
%
\label{analyticMultiplet}
}
\end{center}
\end{table}

\begin{table}[t]
\begin{center}
\begin{tabular}{ccccr}
\hline\hline
$n$     &       $\Lambda$       &       $S$     &       degeneracy      &       $E_{\rm mult}$[meV]\\
\hline
0       &       $A_g$   &       0       &       1       &       0       \\
\hline
1       &       $H_u$   &       1/2     &       10      &       0       \\
\hline
2       &       $T_{2g}$        &       1       &       9       &       -59     \\
        &       $T_{1g}$        &       1       &       9       &       -59     \\
        &       $G_g$   &       1       &       12      &       -11     \\
        &       $G_g$   &       0       &       4       &       17      \\
        &       $H_g$   &       0       &       5       &       46      \\
        &       $H_g$   &       0       &       5       &       115     \\
        &       $A_g$   &       0       &       1       &       318     \\
\hline
3       &       $T_{2u}$        &       3/2     &       12      &       -138    \\
        &       $T_{1u}$        &       3/2     &       12      &       -138    \\
        &       $G_u$   &       3/2     &       16      &       -91     \\
        &       $T_{2u}$        &       1/2     &       6       &       -39     \\
        &       $T_{1u}$        &       1/2     &       6       &       -38     \\
        &       $H_u$   &       1/2     &       10      &       -9      \\
        &       $H_u$   &       1/2     &       10      &       6       \\
        &       $G_u$   &       1/2     &       8       &       23      \\
        &       $T_{2u}$        &       1/2     &       6       &       71      \\
        &       $T_{1u}$        &       1/2     &       6       &       71      \\
        &       $G_u$   &       1/2     &       8       &       96      \\
        &       $H_u$   &       1/2     &       10      &       99      \\
        &       $H_u$   &       1/2     &       10      &       247     \\
\hline
4       &       $H_g$   &       2       &       25      &       -238    \\
        &       $G_g$   &       1       &       12      &       -106    \\
        &       $H_g$   &       1       &       15      &       -94     \\
        &       $G_g$   &       1       &       12      &       -61     \\
        &       $T_{2g}$        &       1       &       9       &       -59     \\
        &       $T_{1g}$        &       1       &       9       &       -59     \\
        &       $G_g$   &       0       &       4       &       -41     \\
        &       $A_g$   &       0       &       1       &       -18     \\
        &       $H_g$   &       1       &       15      &       8       \\
        &       $H_g$   &       0       &       5       &       8       \\
        &       $T_{2g}$        &       1       &       9       &       9       \\
        &       $T_{1g}$        &       1       &       9       &       9       \\
        &       $H_g$   &       1       &       15      &       19      \\
        &       $H_g$   &       0       &       5       &       34      \\
        &       $T_{2g}$        &       0       &       3       &       51      \\
        &       $T_{1g}$        &       0       &       3       &       51      \\
        &       $A_g$   &       0       &       1       &       65      \\
        &       $G_g$   &       0       &       4       &       113     \\
        &       $G_g$   &       1       &       12      &       120     \\
        &       $H_g$   &       0       &       5       &       136     \\
        &       $T_{2g}$        &       1       &       9       &       137     \\
        &       $T_{1g}$        &       1       &       9       &       137     \\
        &       $G_g$   &       0       &       4       &       146     \\
        &       $H_g$   &       0       &       5       &       186     \\
        &       $G_g$   &       0       &       4       &       256     \\
        &       $H_g$   &       0       &       5       &       280     \\
        &       $A_g$   &       0       &       1       &       494     \\
\hline\hline
\end{tabular}
\begin{tabular}{ccccr}
\hline\hline
$n$     &       $\Lambda$       &       $S$     &       degeneracy      &       $E_{\rm mult}$[meV]\\
\hline
5       &       $A_u$   &       5/2     &       6       &       -397    \\
        &       $H_u$   &       3/2     &       20      &       -195    \\
        &       $H_u$   &       3/2     &       20      &       -125    \\
        &       $G_u$   &       3/2     &       16      &       -97     \\
        &       $H_u$   &       1/2     &       10      &       -82     \\
        &       $G_u$   &       1/2     &       8       &       -69     \\
        &       $G_u$   &       3/2     &       16      &       -68     \\
        &       $A_u$   &       1/2     &       2       &       -62     \\
        &       $T_{2u}$        &       3/2     &       12      &       -21     \\
        &       $T_{1u}$        &       3/2     &       12      &       -20     \\
        &       $G_u$   &       1/2     &       8       &       -14     \\
        &       $H_u$   &       1/2     &       10      &       3       \\
        &       $H_u$   &       1/2     &       10      &       9       \\
        &       $T_{2u}$        &       1/2     &       6       &       47      \\
        &       $T_{1u}$        &       1/2     &       6       &       47      \\
        &       $T_{1u}$        &       1/2     &       6       &       54      \\
        &       $T_{2u}$        &       1/2     &       6       &       55      \\
        &       $H_u$   &       1/2     &       10      &       55      \\
        &       $H_u$   &       1/2     &       10      &       79      \\
        &       $G_u$   &       1/2     &       8       &       101     \\
        &       $A_u$   &       1/2     &       2       &       118     \\
        &       $G_u$   &       1/2     &       8       &       156     \\
        &       $G_u$   &       1/2     &       8       &       160     \\
        &       $T_{2u}$        &       1/2     &       6       &       175     \\
        &       $T_{1u}$        &       1/2     &       6       &       175     \\
        &       $H_u$   &       1/2     &       10      &       185     \\
        &       $H_u$   &       1/2     &       10      &       334     \\
\hline\hline
\end{tabular}
\caption{
The Coulomb multiplet energies for the Coulomb parameters obtained for
C$_{60}^{n+}$.
The states are sorted by increasing energy.
The non particle-hole symmetric contribution
$\left[\epsilon n + U n(n-1)/2\right]$ is not included in these numbers.
\label{numericMultiplet}
}
\end{center}
\end{table}

\begin{table}[t]
\begin{center}
\begin{tabular}{ccccr}
\hline\hline
$n$     &       $\Lambda$       &       $S$     &       degeneracy      &       $E_{\rm mult}$[meV]\\
\hline
0       &       $A_g$   &       0       &       1       &       0       \\
\hline
1       &       $H_u$   &       1/2     &       10      &       0       \\
\hline
2       &       $T_{1g}$        &       1       &       9       &       -33     \\
        &       $H_g$   &       0       &       5       &       -21     \\
        &       $T_{2g}$        &       1       &       9       &       -1      \\
        &       $G_g$   &       0       &       4       &       3       \\
        &       $A_g$   &       0       &       1       &       15      \\
        &       $G_g$   &       1       &       12      &       21      \\
        &       $H_g$   &       0       &       5       &       27      \\
\hline
3       &       $H_u$   &       1/2     &       10      &       -50     \\
        &       $G_u$   &       1/2     &       8       &       -45     \\
        &       $T_{1u}$        &       3/2     &       12      &       -36     \\
        &       $T_{2u}$        &       1/2     &       6       &       -18     \\
        &       $T_{2u}$        &       3/2     &       12      &       -5      \\
        &       $G_u$   &       1/2     &       8       &       -4\\ 
        &       $H_u$   &       1/2     &       10      &       -4\\ 
        &       $T_{2u}$        &       1/2     &       6       &       9       \\
        &       $T_{1u}$        &       1/2     &       6       &       9       \\
        &       $T_{1u}$        &       1/2     &       6       &       14      \\
        &       $G_u$   &       3/2     &       16      &       17      \\
        &       $H_u$   &       1/2     &       10      &       49      \\
        &       $H_u$   &       1/2     &       10      &       58      \\
\hline
4       &       $A_g$   &       0       &       1       &       -90     \\
        &       $H_g$   &       0       &       5       &       -79     \\
        &       $T_{1g}$        &       1       &       9       &       -54     \\
        &       $T_{2g}$        &       1       &       9       &       -47     \\
        &       $H_g$   &       0       &       5       &       -41     \\
        &       $G_g$   &       0       &       4       &       -39     \\
        &       $G_g$   &       1       &       12      &       -38     \\
        &       $H_g$   &       1       &       15      &       -31     \\
        &       $G_g$   &       0       &       4       &       -18     \\
        &       $H_g$   &       1       &       15      &       -13     \\
        &       $H_g$   &       2       &       25      &       -11     \\
        &       $T_{1g}$        &       1       &       9       &       -2      \\
        &       $T_{2g}$        &       1       &       9       &       2       \\
        &       $H_g$   &       0       &       5       &       3       \\
        &       $T_{2g}$        &       0       &       3       &       6       \\
        &       $G_g$   &       1       &       12      &       6       \\
        &       $A_g$   &       0       &       1       &       10      \\
        &       $G_g$   &       0       &       4       &       21      \\
        &       $H_g$   &       0       &       5       &       23      \\
        &       $H_g$   &       1       &       15      &       31      \\
        &       $T_{1g}$        &       0       &       3       &       38      \\
        &       $T_{2g}$        &       1       &       9       &       43      \\
        &       $G_g$   &       1       &       12      &       50      \\
        &       $T_{1g}$        &       1       &       9       &       54      \\
        &       $H_g$   &       0       &       5       &       63      \\
        &       $G_g$   &       0       &       4       &       101     \\
        &       $A_g$   &       0       &       1       &       116     \\
\hline\hline
\end{tabular}
\begin{tabular}{ccccr}
\hline\hline
$n$     &       $\Lambda$       &       $S$     &       degeneracy      &       $E_{\rm mult}$[meV]\\
\hline
5       &       $T_{1u}$        &       1/2     &       6       &       -84     \\
        &       $H_u$   &       1/2     &       10      &       -74     \\
        &       $G_u$   &       1/2     &       8       &       -49     \\
        &       $G_u$   &       1/2     &       8       &       -45     \\
        &       $A_u$   &       1/2     &       2       &       -39     \\
        &       $H_u$   &       1/2     &       10      &       -31     \\
        &       $H_u$   &       3/2     &       20      &       -30     \\
        &       $G_u$   &       3/2     &       16      &       -24     \\
        &       $T_{2u}$        &       1/2     &       6       &       -19     \\
        &       $A_u$   &       5/2     &       6       &       -18     \\
        &       $H_u$   &       1/2     &       10      &       -17     \\
        &       $H_u$   &       1/2     &       10      &       -8      \\
        &       $G_u$   &       3/2     &       16      &       -7      \\
        &       $T_{2u}$        &       3/2     &       12      &       -2      \\
        &       $A_u$   &       1/2     &       2       &       2       \\
        &       $T_{2u}$        &       1/2     &       6       &       4       \\
        &       $H_u$   &       1/2     &       10      &       8       \\
        &       $T_{1u}$        &       1/2     &       6       &       12      \\
        &       $G_u$   &       1/2     &       8       &       17      \\
        &       $H_u$   &       3/2     &       20      &       17      \\
        &       $T_{1u}$        &       3/2     &       12      &       29      \\
        &       $G_u$   &       1/2     &       8       &       33      \\
        &       $H_u$   &       1/2     &       10      &       41      \\
        &       $G_u$   &       1/2     &       8       &       48      \\
        &       $T_{1u}$        &       1/2     &       6       &       81      \\
        &       $T_{2u}$        &       1/2     &       6       &       88      \\
        &       $H_u$   &       1/2     &       10      &       91      \\
\hline\hline
\end{tabular}
\end{center}
\caption{
The C$_{60}^{n+}$ multiplet energies for the effective Coulomb parameters,
including the e-ph coupling in the anti-adiabatic (weak coupling) limit.
The states are sorted by increasing energy.
The non particle-hole symmetric contribution
$\left[\epsilon n + U^{\rm tot} n(n-1)/2\right]$ is not included
in these results.
\label{effectiveMultiplet}}
\end{table}

\begin{table}[ht]
\begin{center}
\begin{tabular}{cc|rr}
\hline
\hline
$n\pm$&$S$ & adiabatic & anti-adiabatic \\
\hline
 2+  &  0  & -129    &  -21 \\
     &  1  & {\bf -142} & {\bf -33} \\
& & \\
 3+  & 1/2 & -168    & {\bf -50} \\
     & 3/2 & {\bf -222} & -36 \\
& & \\
 4+  &  0  & -200 & {\bf -90} \\
     &  1  & -211 & -54 \\
     &  2  & {\bf -308} & -11 \\
& & \\
 5+  & 1/2 & -203 & {\bf -84} \\ 
     & 3/2 & -256 & -30 \\
     & 5/2 & {\bf -397} & -18 \\
\hline
$2-$ &  0  &  {\bf -92} & {\bf -100}\\
     &  1  &  -71 &  -25\\
& & \\
$3-$ & 1/2 &  -85 & {\bf -50}\\
     & 3/2 &  {\bf -97} & +75 \\
\hline
\hline
\end{tabular}
\end{center}
\caption{
%
The energy (in meV) of the lowest state for each $n$ and $S$, including the
e-e and e-ph contributions from $\hat{H}_{\rm vib} + \hat{H}_{\rm e-vib} +
\hat{H}_{\rm e-e}$ (but excluding the $\left[\epsilon n + U
n(n-1)/2\right]$ term), for $h_u$ holes (upper panel) and $t_{1u}$
electrons (lower).
First column: the phonons are treated in the adiabatic (strong-coupling)
approximation, by full relaxation of the phonon modes to the optimal
classical JT distortion for each $n$ and $S$.
Second column: the lowest energy for each $n$ and $S$ is reported from
Table~\ref{effectiveMultiplet}.
The anti-adiabatic energy lowerings are smaller than the adiabatic ones
because of a larger cancellation of e-e and e-ph contributions.
\label{adiabatic_energies}
}
\end{table}


\end{document}